\documentclass[10pt]{article}

\usepackage{fancyhdr}
\usepackage{extramarks}
\usepackage{amsmath}
\usepackage{amsthm}
\usepackage{amsfonts}
\usepackage{siunitx}
\usepackage{tikz}
\usepackage[plain]{algorithm}
\usepackage{algpseudocode}
\usepackage{multirow}
\usepackage{booktabs}
\usepackage{graphicx}
\usepackage{subfigure}
\usepackage[colorlinks,linkcolor=black,anchorcolor=black,citecolor=black,urlcolor=blue]{hyperref}
\usepackage{amsmath,bm}
\usepackage{booktabs}
\usepackage{mathtools}
\usepackage{amssymb}
\usepackage{caption}
\usepackage{capt-of}
\usepackage{mciteplus}
\usepackage{cite}
\usepackage{mathrsfs}
\usepackage[title,titletoc,toc]{appendix}
\usepackage{xr}
\usepackage{parskip}
\usetikzlibrary{automata,positioning}

\renewcommand{\part}[1]{\textbf{\large Part \Alph{partCounter}}\stepcounter{partCounter}\\}

\begin{document}

\title{Mutations strengthened SARS-CoV-2 infectivity  }
\author{ Jiahui Chen$^1$, Rui Wang$^1$, Menglun Wang$^1$, and Guo-Wei Wei$^{1,2,3}$\footnote{
		Corresponding author.		Email: wei@math.msu.edu} \\
$^1$ Department of Mathematics, \\
Michigan State University, MI 48824, USA.\\
$^2$ Department of Electrical and Computer Engineering,\\
Michigan State University, MI 48824, USA. \\
$^3$ Department of Biochemistry and Molecular Biology,\\
Michigan State University, MI 48824, USA. \\
}
\date{\today} 

\maketitle

\begin{abstract}
Severe acute respiratory syndrome coronavirus 2 (SARS-CoV-2) infectivity is a major concern in coronavirus disease 2019 (COVID-19)  prevention and economic reopening.  However,  rigorous determination of SARS-COV-2 infectivity is essentially impossible owing to its continuous evolution with over 13752 single nucleotide polymorphisms (SNP) variants in six different subtypes.  We develop an advanced machine learning algorithm based on the algebraic topology to quantitatively evaluate the binding affinity changes of SARS-CoV-2 spike glycoprotein (S protein) and host angiotensin-converting enzyme 2 (ACE2) receptor following the mutations.  Based on mutation-induced binding affinity changes, we reveal that five out of six SARS-CoV-2 subtypes have become either moderately or slightly more infectious, while one subtype has weakened its infectivity. We find that SARS-CoV-2 is slightly more infectious than SARS-CoV according to computed S protein-ACE2 binding affinity changes.  Based on a systematic evaluation of all possible 3686 future mutations on the S protein receptor-binding domain (RBD), we show that most likely future mutations will make SARS-CoV-2 more infectious. Combining sequence alignment,  probability analysis, and binding affinity calculation, we predict that a few residues on the receptor-binding motif (RBM), i.e.,  452, 489, 500, 501, and 505,  have very high chances to mutate into significantly more infectious COVID-19 strains. 

\end{abstract}

\pagenumbering{roman}
\begin{verbatim}
\end{verbatim}
  \newpage

{\setcounter{tocdepth}{4} \tableofcontents}

  \setcounter{page}{1}
 \renewcommand{\thepage}{{\arabic{page}}}
\section{Introduction}
In December 2019, an outbreak of pneumonia due to coronavirus disease 2019 (COVID-19) was initially detected in Wuhan, China \cite{huang2020clinical}, due to severe acute respiratory syndrome coronavirus 2 (SARS-CoV-2). It has now spread globally via travelers and breached the boundaries of 213 countries and territories, leading to more than 5.4 million infection cases and 343,000 deaths as of   May 23, 2020.    In the past two decades, there have been three major zoonotic disease outbreaks of betacoronaviruses: SARS-CoV in 2002, Middle East respiratory syndrome coronavirus (MERS-CoV) in 2012, and SARS-CoV-2 in 2019. Similar to SARS-CoV and MERS-CoV, SARS-CoV-2 infections were observed in hospital personnel and family clusters in the early stages of the outbreak \cite{lu2020genomic, chan2020familial, gralinski2020return}. Unfortunately,  there is no specific antivirus drugs nor effective vaccines developed to moderate this outbreak at present.

SARS-CoV-2 is an enveloped non-segmented positive-sense RNA virus and belongs to the betacoronavirus genus. Although the origin of SARS-CoV-2 remains elusive, it has undergone thousands of recorded single mutations compared to the reference genome collected on January 5, 2020 \cite{wu2020new,wang2020decoding}.  Because of the lack of proofreading ability in RNA polymerases, RNA viruses are prone to random mutations. Human immune system intervention introduces viral mutations too. However, rapid global spread and transmission of COVID-19 provide the virus with substantial opportunities for the natural selection of rare-acted but favorable mutations.
Therefore, although some viral mutations are benign, many mutations strengthen viral survival capability. It is of paramount importance to understand SARS-CoV-2 infectivity changes following the existing mutations and predict the future infection tendency.  
 
It is well known that like SARS-CoV, SARS-CoV-2 enters host cells through the interaction of spike glycoprotein (S protein) and host angiotensin-converting enzyme 2 (ACE2) receptor  \cite{ li2005structure,hoffmann2020sars,walls2020structure}.  In both SARS-CoV and SARS-CoV-2, the S protein receptor-binding domain (RBD) is recognized as one of two subunits, i.e., S1 and S2,  to bind directly to the  ACE2.  Although the SARS-CoV-2 S protein harbors a furin cleavage site at the boundary between the S1/S2 subunits  \cite{wu2020new}, lessons learned from SARS-CoV are important in formulating hypotheses about SARS-CoV-2, as well as the receptor recognition when studying the host range, cross-species transmission, and pathogenesis of SARS-CoV-2. In the studies of SARS-CoV, epidemiologic and biochemical studies show that the infectivity of different SARS-CoV strains in host cells is proportional to the binding affinity between the RBD of each strain and the ACE2 expressed by the host cell \cite{li2005bats, qu2005identification, song2005cross,hoffmann2020sars,walls2020structure}. Therefore, the assessment of binding affinity changes following mutations is vital for the understanding of SARS-CoV-2 infectivity evolution.   

It is very challenging to rigorously measure the relative viral infectivity of two viruses by experiments. There is a discrepancy in the literature about the relative S protein-ACE2 binding affinities of  SARS-CoV and  SARS-CoV-2. Wrapp {\it et al} and Shang {\it et al} reported that SARS-CoV-2  has a higher binding affinity than SARS-CoV does \cite{wrapp2020cryo, shang2020structural}, whereas Walls {\it et al} argued that   SARS-CoV-2 and SARS-CoV bind with similar affinities to ACE2. \cite{walls2020structure} 

The first SARS-CoV-2 genome reported on January 5, 2020 \cite{wu2020new} has about 80\% sequence identity with that of SARS-CoV.  However, compared with SARS-CoV, SARS-CoV-2 S protein has 301 mutations over its 1255 residues.  Their sequence identity is only 76\%. Among 301 mutations on SARS-CoV-2 S protein, 50 were on the RBD, which has a total of 194 residues, suggesting that the RBD is subject to more mutations.  Our recent studies using over 13,000 genome samples show that SARS-CoV-2 S protein is among the most non-conservative ones in its genome \cite{wang2020decoding}. Since early January 2020,  hundreds of new mutations were found on different residue positions of SARS-CoV-2 S protein. Many of them are located on the RBD \cite{wang2020decoding}. The existence of so many different S protein mutations indicates that there are many different SARS-CoV-2 subtypes that might have very different infectivities. Obviously, the relatively high mutation rate at the RBD poses a real threat to the occurrence of future SARS-CoV-2 strains that might be more infectious than the current SARS-CoV-2.  
  
Due to the continuous evolution of SARS-CoV-2, the experimental measurement of SARS-CoV-2 infectivity is extremely difficult if it is not entirely impossible. The computational estimation of mutation-induced protein-protein binding affinity changes is an important approach for understanding the impact of mutations on protein-protein interactions (PPIs). There are many standard databases available, including the AB-Bind database of mutation-induced antibody-antigen complex binding free energy changes \cite{sirin2016ab} and  SKEMPI  for protein-protein binding affinity changes upon mutation ($\Delta\Delta G$) \cite{moal2012skempi}. These databases have been used as a benchmark for evaluating the predictive power of various computational methods \cite{ 
 schymkowitz2005foldx,
 pires2016mcsm}. To simplify the structural complexity of protein-protein complexes, we have recently introduced element-specific and site-specific persistent homology, a new branch of algebraic topology, to embed molecular mechanisms into topological invariants \cite{wang2020topology}. This approach is paired with a new deep learning algorithm called NetTree, to combined convolutional neural networks and gradient boosting trees. The resulting method, called TopNetTree, was about 22\% better than the previous best result for the AB-Bind dataset and significantly outperformed the state-of-the-art in the literature on the SKEMPI database \cite{wang2020topology}. 

The objective of this work is three-fold. First, we train the TopNetTree on a set of 8338 PPI data to analyze the impacts of existing S protein RBD mutations on the binding affinity of the S protein and the ACE2. Since different SARS-CoV-2 subtypes have different mutation patterns, it is important to understand their mutation impacts accordingly.  We carry our analysis based on existing six mutation clusters  \cite{wang2020decoding}, though more specific analysis can be easily done as well. Additionally, it is also extremely important to know whether future SARS-CoV-2 subtypes would pose an imminent danger to public health. To this end, we have conducted a systematic screening of all possible 3,686 future mutations on all 194 residues (residue IDs from 333 to 526 on S protein) on the RBD. We classify these mutations into three categories: the most likely ones which would happen by a single mutation at any one of three constitutive nucleotides; the likely mutations which would occur via two concurrent mutations at three constitutive nucleotides; and the unlikely mutations which would produce through three concurrent mutations at all of three constitutive nucleotides. Finally, we analyze how 50 mutations on the RBD of the SARS-CoV-2 S protein with respect to  SARS-CoV have changed its infectivity. 

\begin{figure}[H]
    \centering
    \captionsetup{margin=0.5cm}
    \includegraphics[scale=0.25]{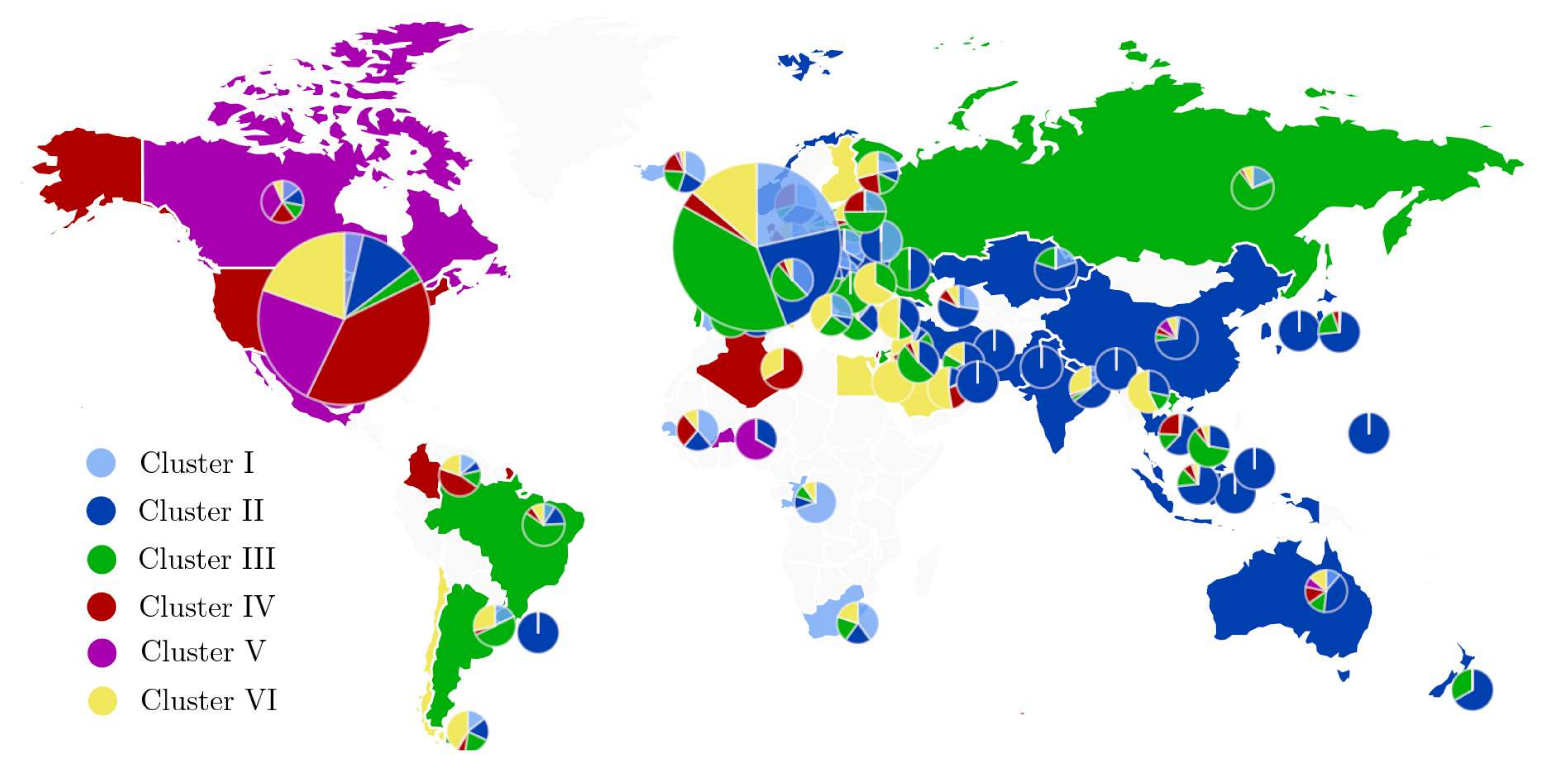}
    \caption{The scatter plot of six distinct clusters in the world. The light blue, dark blue, green, red, pink, and yellow represent Cluster I, Cluster II, Cluster III, Cluster IV, Cluster V, and Cluster VI, respectively. The size of the pie charts corresponds to the number of samples. The color of the dominated cluster decides the base color of each country.}
    \label{fig:World_Cluster}
\end{figure}

\section{Results}
 \begin{table}[H]
    \centering
    \setlength\tabcolsep{5pt}
    \captionsetup{margin=0.9cm}
    \caption{The cluster distributions of samples ($N_{\rm NS}$) and total mutation counts ($N_ {\rm TF}$) for  17 countries  \cite{wang2020decoding}.}
    \begin{tabular}{lcc|cc|cc|cc|cc|cc}
    \hline
      &   \multicolumn{2}{c|}{Cluster I}    & \multicolumn{2}{c|}{Cluster II}    & \multicolumn{2}{c|}{Cluster III}    & \multicolumn{2}{c|}{Cluster IV}   & \multicolumn{2}{c|}{Cluster V} & \multicolumn{2}{c}{Cluster VI}    \\ \hline
          Country                &   $N_{\rm NS}$ &$N_{\rm TF}$  & $N_{\rm NS}$ &$N_{\rm TF}$  & $N_{\rm NS}$ &$N_{\rm TF}$  & $N_{\rm NS}$ &$N_{\rm TF}$  &$N_{\rm NS}$ &$N_{\rm TF}$ &$N_{\rm NS}$ &$N {\rm TF}$\\  \hline
     US             & 146&829   & 450&2948   & 126&1171    & 1600&11963   & 946&6766  & 793&8539  \\
     CA             & 17&95     & 16&79      & 14&119      & 23&154       & 40&240    & 8&76   \\
     AU             & 99&555    & 357&3533   & 118&1020    & 115&827      & 69&469    & 122&1423    \\
 UK             & 852&5073  & 908&6031   & 1545&14549  & 124&961      & 3&15      & 544&5175        \\
 IS             & 145&868   & 89&474     & 89&870      & 70&472       & 15&127    & 17&145 \\
   ES             & 111&677   & 84&555     & 25&217      & 7&59         & 2&6       & 35&307  \\
	 CN             & 2&8       & 192&910    & 1&13        & 1&7          & 25&58     & 8&69    \\
     DE             & 20&100    & 20&97      & 38&324      & 41&274       & 0&0       & 10&67     \\ 
     FR             & 64&385    & 14&55      & 12&105      & 98&49        & 0&0       & 46&446  \\
		IN             & 33&184    & 114&1145   & 9&84        & 5&40         & 0&0       & 69&753  \\
     RU             & 18&95     & 1&2        & 68&583      & 3&21         & 0&0       & 7&63   \\
     BE             & 111&647   & 39&151     & 91&815      & 16&114       & 0&0       & 68&599  \\
     SA             & 1&5       & 9&61       & 1&7         & 16&110       & 0&0       & 30&254  \\
	  IT             & 18&93     & 5&110      & 17&183      & 0&0          & 0&0       & 26&426 \\
     JP             & 1&6       & 68&200     & 20&169      & 5&32         & 0&0       & 0&0  \\
		     TR             & 1&4       & 27&244     & 9&72        & 0&0          & 0&0       & 36&295  \\
     KR             & 0&0       & 28&167     & 0&0         & 0&0          & 0&0       & 0&0  \\
     \hline
    \end{tabular}
    \label{table:World}
\end{table}

 \subsection{Impacts of existing RBD mutations}

 \subsubsection{Global analysis}
To investigate the influences of existing S protein RBD mutations on binding affinity (BA) of S protein and ACE2, the 13752 complete  SARS-CoV-2 genome samples deposited at GISAID \cite{shu2017gisaid} are compared with the first genome sequence of SARS-CoV-2 collected on January 5, 2020 \cite{wu2020new}.  The resulting 7525 single mutations are found in six distinct clusters as shown in Table \ref{table:World}\cite{wang2020decoding} and Figure \ref{fig:World_Cluster}.  There are 434 existing non-degenerated mutations on SARS-CoV-2 S protein. Among them, 55 mutations occurred on the RBD which are relevant to the binding of SARS-CoV-2 S protein and ACE2. Furthermore,  28 out of 55 mutations are on the receptor-binding motif (RBM), i.e., the region of RBD that is in direct contact with the ACE2. 

We examine the free energy changes following the existing site-specific mutations. Our studies are based on the X-ray crystal structure of SARS-CoV-2 S protein and ACE2 (PDB 6M0J)  \cite{lan2020structure} (see Fig.~\ref{fig:NetTree}), whose S protein gene sequence is consistent with that of the reference  SARS-CoV-2 \cite{wu2020new}. The BA change following mutation ($\Delta\Delta G$) is defined as the subtraction of the BA of the mutant from the BA of wild type, $\Delta\Delta G=\Delta G_{\rm W} - \Delta G_{\rm M}$ where $\Delta G_{\rm W}$ is BA of the wild type and $\Delta G_{\rm M}$ is BA of mutant type.  Therefore, a positive BA change means that the mutation increases affinities, making the mutant more stable and more infective.

\begin{figure}
	\includegraphics[width=1\linewidth]{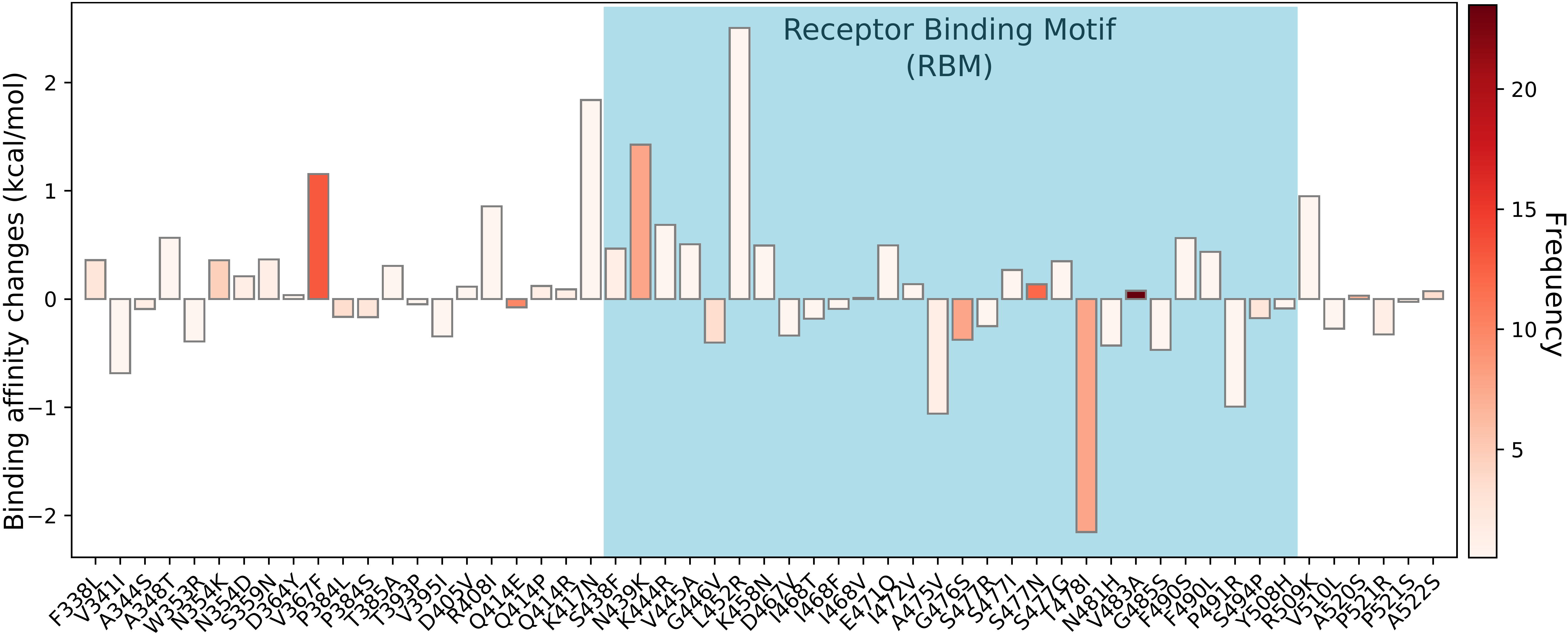}
	\caption{Overall binding affinity changes $\Delta\Delta G$ on the receptor-binding domain (RBD).  The blue color region marks the binding affinity changes on the receptor-binding motif (RBM). The height of each bar indicates the predicted $\Delta\Delta G$. The color indicates the occurrence frequency in the GISAID genome dataset. }
	\label{fig:overall}
\end{figure}

We present the overall BA changes $\Delta\Delta G$ of SARS-CoV-2 S protein RBD in Figure~\ref{fig:overall}. Most mutations have a small number of changes in their binding affinities, while some of them have large changes. There are 56\% mutations on RBD having positive BA changes (i.e., 31 over 55) including  V367F and V483A, which have the most frequencies, 13 and 23 respectively. This statistic implies that the evolution of SARS-CoV-2 is mostly driven by selection and COVID-19 evolves toward more infectious.  It is noted that many mutations on the RBM,  such as N439K, L452R, and T478I, have significant free energy changes. The mutations on the RBM take 51\% (28 over 55) of all mutations on the RBD, which potentially increases the complexity of antiviral drug and vaccine development. This global analysis indicates that mutations on RBD strengthen the binding of S protein and ACE2, leading to more infectious SARS-CoV-2. 

The SARS-CoV-2 genotypes are clustered into six clusters or subtypes based on their single nucleotide polymorphism (SNP) variants \cite{wang2020decoding}. Accordingly, a more detailed analysis of mutation impacts on the  BA changes can be carried out on each cluster, which reveals the diversity of COVID-19   infection rates and provides evidence for transmission pathways and spread dynamics across the world.

It is worth noting that residue 414 has three mutations, Q414P, Q414E, and Q414R, due to mutations at two adjacent nucleotides 22802 and 22803: 22803A$>$C, Q414P;  22802C$>$G, Q414E; and 22803A$>$G, Q414R. At the protein level, some or all of these mutations show up in different clusters. Similarly, residues 354 and 521 have two existing mutations. 

\subsubsection{Cluster I analysis}

\begin{figure}
	\includegraphics[width=1\linewidth]{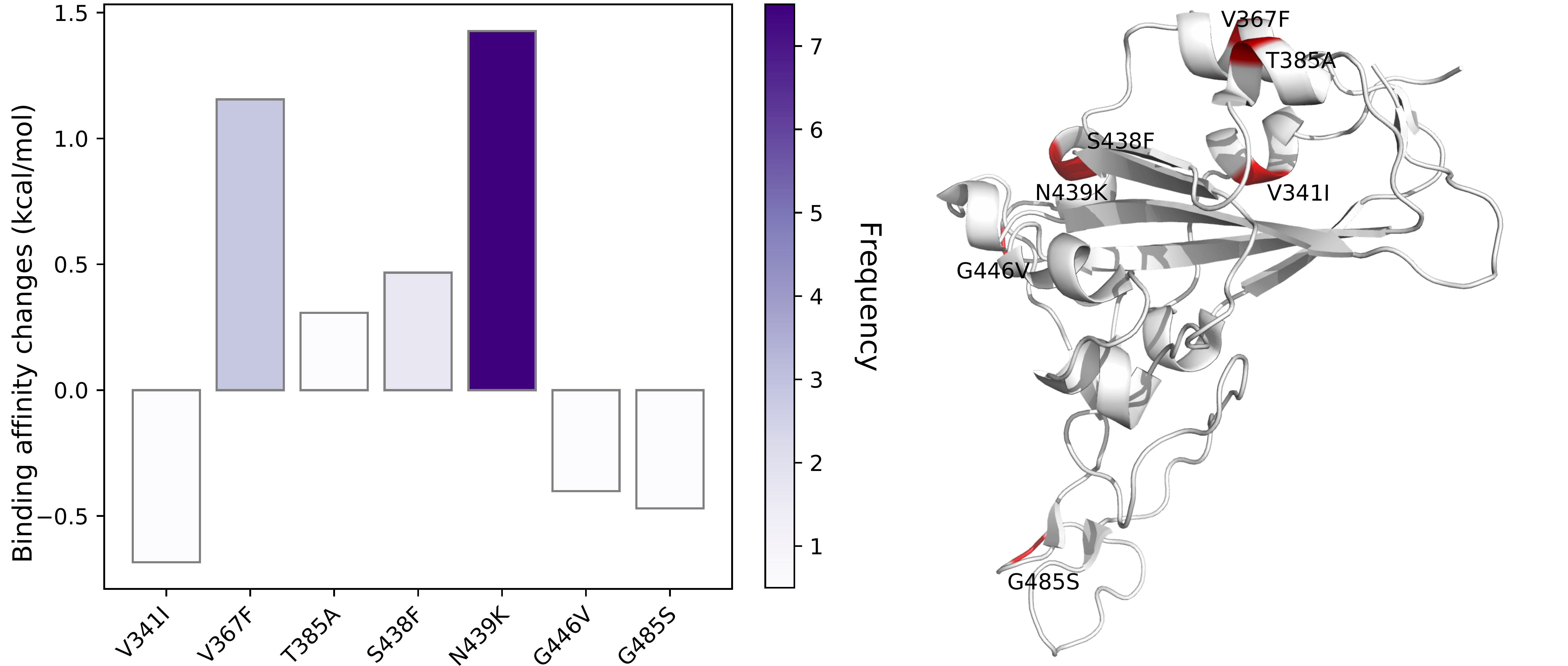}
	\caption{Cluster I. Left: Binding affinity changes $\Delta\Delta G$  induced by   mutations in Cluster I. Right: mutation locations on the SARS-CoV-2 S protein RBD.}
	\label{fig:c1}
\end{figure}

 Figure~\ref{fig:c1} depicts the binding affinity changes $\Delta\Delta G$ of Cluster I. Four out of sixth mutations have positive binding affinity changes which indicate increasing infectivity on Cluster 1. Particularly,  mutation N439K, which has a higher frequency and a larger free energy amplitude than the rest, attained a very positive binding affinity change to increase the affinity between S protein and ACE2.  Therefore, Cluster I has a moderate increase in its infectivity. Cluster I  is associated with COVID-19 in most countries except for Japan and South Korea  \cite{wang2020decoding}.

 \subsubsection{Cluster II analysis}
\begin{figure}
	\includegraphics[width=1\linewidth]{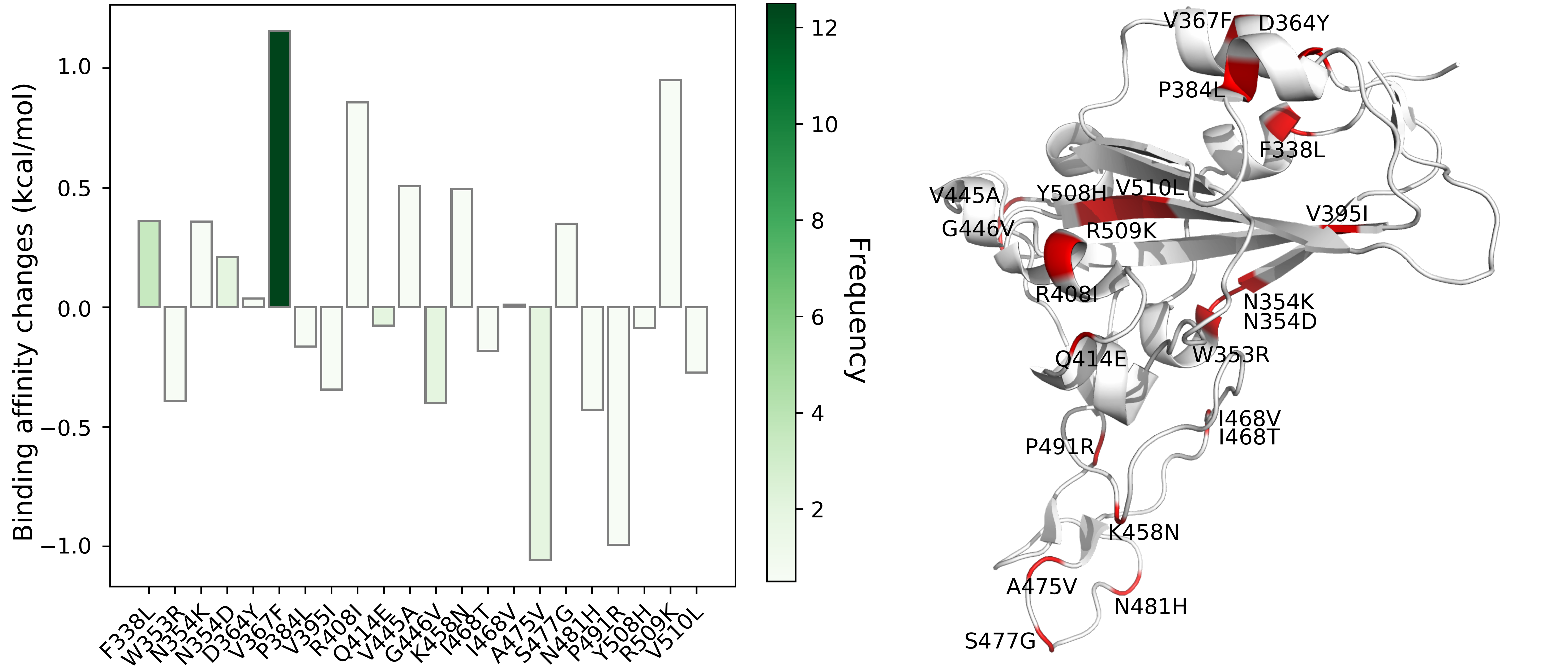}
	\caption{Cluster II. Left: Binding affinity changes $\Delta\Delta G$ induced by  mutations in Cluster II. Right: mutation locations on the SARS-CoV-2 S protein RBD.}
	\label{fig:c2}
\end{figure}

 Figure~\ref{fig:c2} illustrates the binding affinity changes following the mutations of Cluster II.  As shown in the figure, there are many mutations on the RBD. However, most mutations are associated with small free energy changes. When only considering the absolute value of BA change greater than 0.5 kcal/mol, five mutations have positive BA changes whereas two mutations have negative BA changes. The mutation D364Y has the highest frequency and the highest free energy change, indicating the increase in infectivity.  Therefore, Cluster II has a minor increase in infectivity.
Note that Cluster II COVID-19 is found in every country that has submitted SARS-CoV-2 genome samples. 

 \subsubsection{Cluster III analysis} 
\begin{figure}
	\includegraphics[width=1\linewidth]{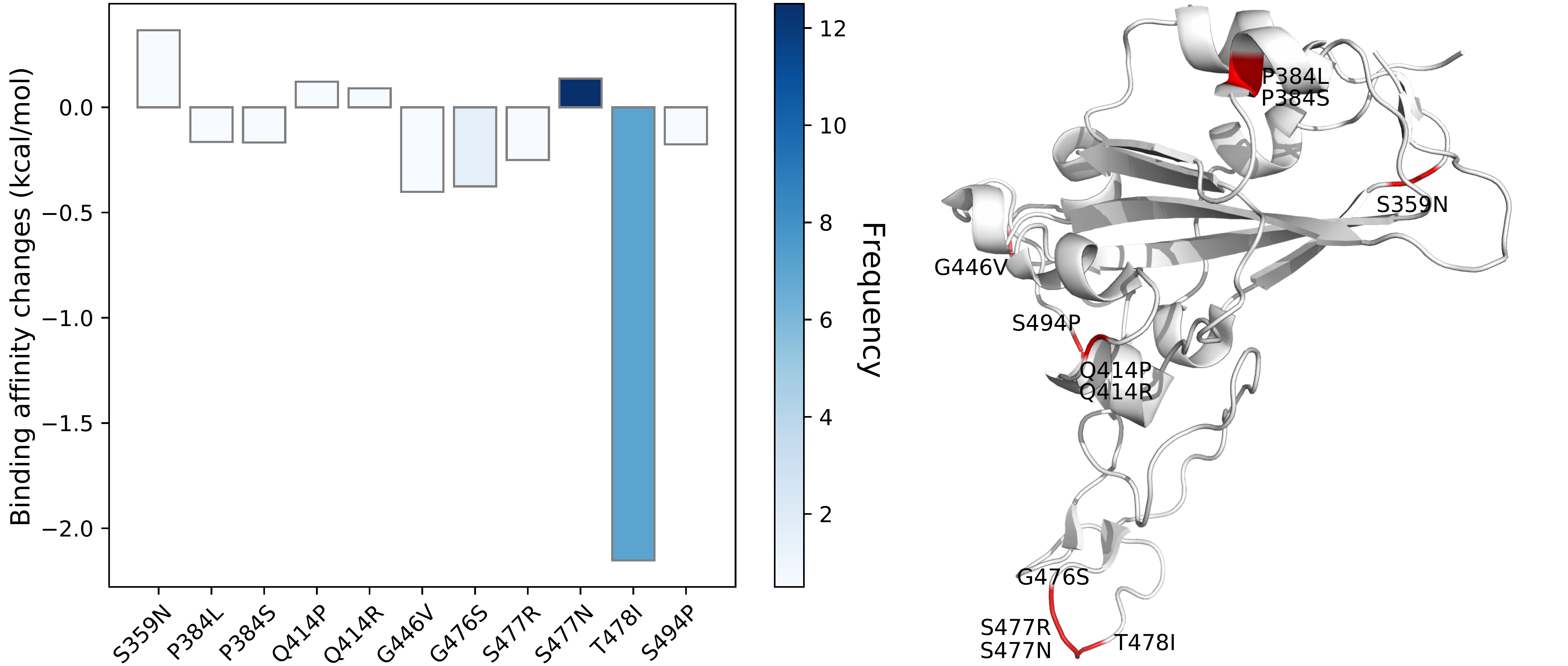}
	\caption{Cluster III. Left: Binding affinity changes $\Delta\Delta G$  induced by   mutations in Cluster III. Right: mutation locations on the SARS-CoV-2 S protein RBD.}
	\label{fig:c3}
\end{figure}

From  Figure~\ref{fig:c3}, a significant decreasing trend of affinity is observed such that the largest change is also the most frequent mutation with a negative BA change while the rest of the mutation-induced BA changes are negligible. Interestingly, the mutation T478I changes from amino acid with polar uncharged side chains, Threonine, to amino acids with hydrophobic side chains, Isoleucine, which significantly decreases affinity between the S protein and the ACE2 receptor. Another observation is that mutations that happened in the same residues have close changes such as P384L and P384S, S477R and S477N, or Q414P and Q414R. Cluster III is only one which has a decreasing trend of binding affinities as SARS-CoV-2 evolving from this cluster. This cluster involves genome samples from all countries except for South Korea. Notably, most samples are submitted by the UK.

 \subsubsection{Cluster IV analysis} 
\begin{figure}
	\includegraphics[width=1\linewidth]{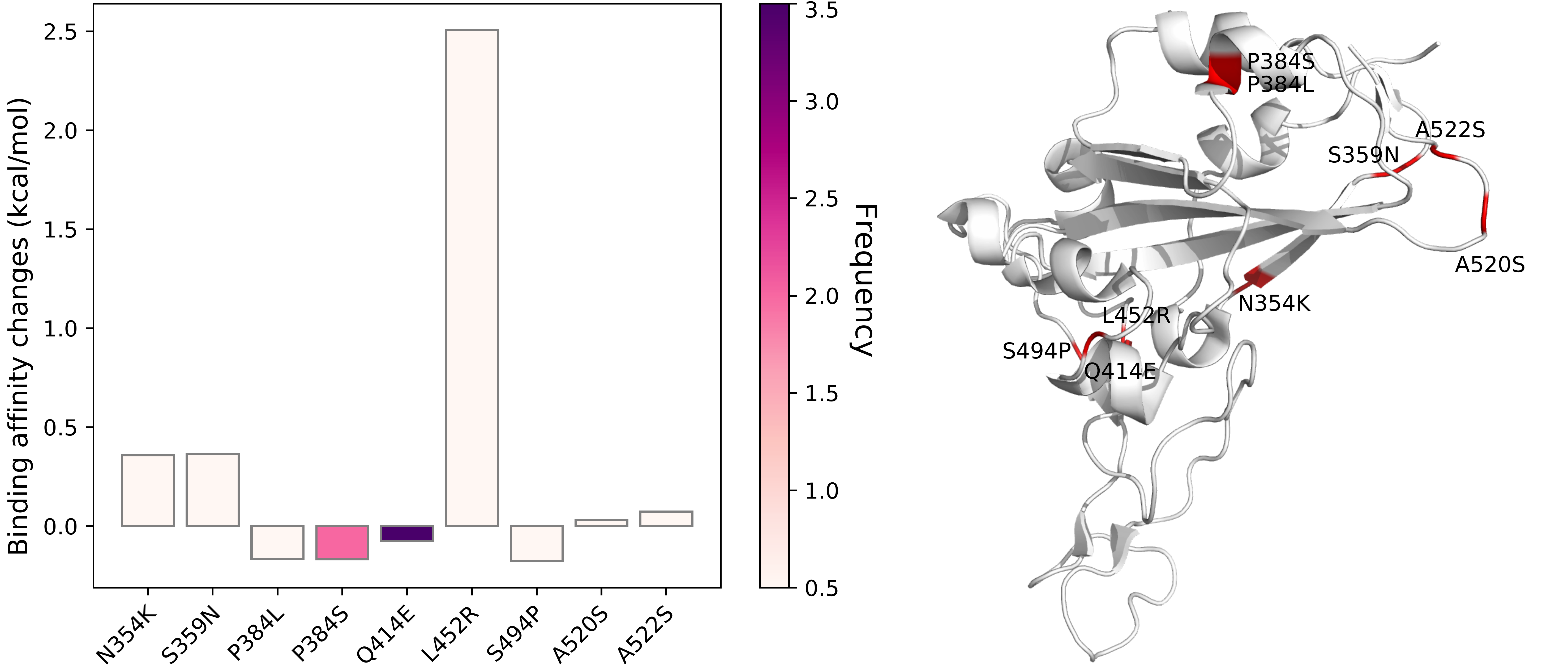}
	\caption{Cluster IV. Left: Binding affinity changes $\Delta\Delta G$  induced by   mutations in Cluster IV. Right: mutation locations on the SARS-CoV-2 S protein RBD.}
	\label{fig:c4}
\end{figure}

Figure~\ref{fig:c4} shows the binding affinity changes of mutations in Cluster IV. Among all mutations in Cluster 4, mutation L452R has the largest free energy change and directly connects the ACE2 receptor. Although the most frequent mutation Q414E has a  negative change, the BA change $\Delta\Delta G$ is -0.055 kcal/mol which is negligible compared with others. The overall trend of this cluster is considered as increasing the COVID-19 infectibility. Note that most genome samples in Cluster IV are submitted by the US. 

\subsubsection{Cluster V analysis}
\begin{figure}
	\includegraphics[width=1\linewidth]{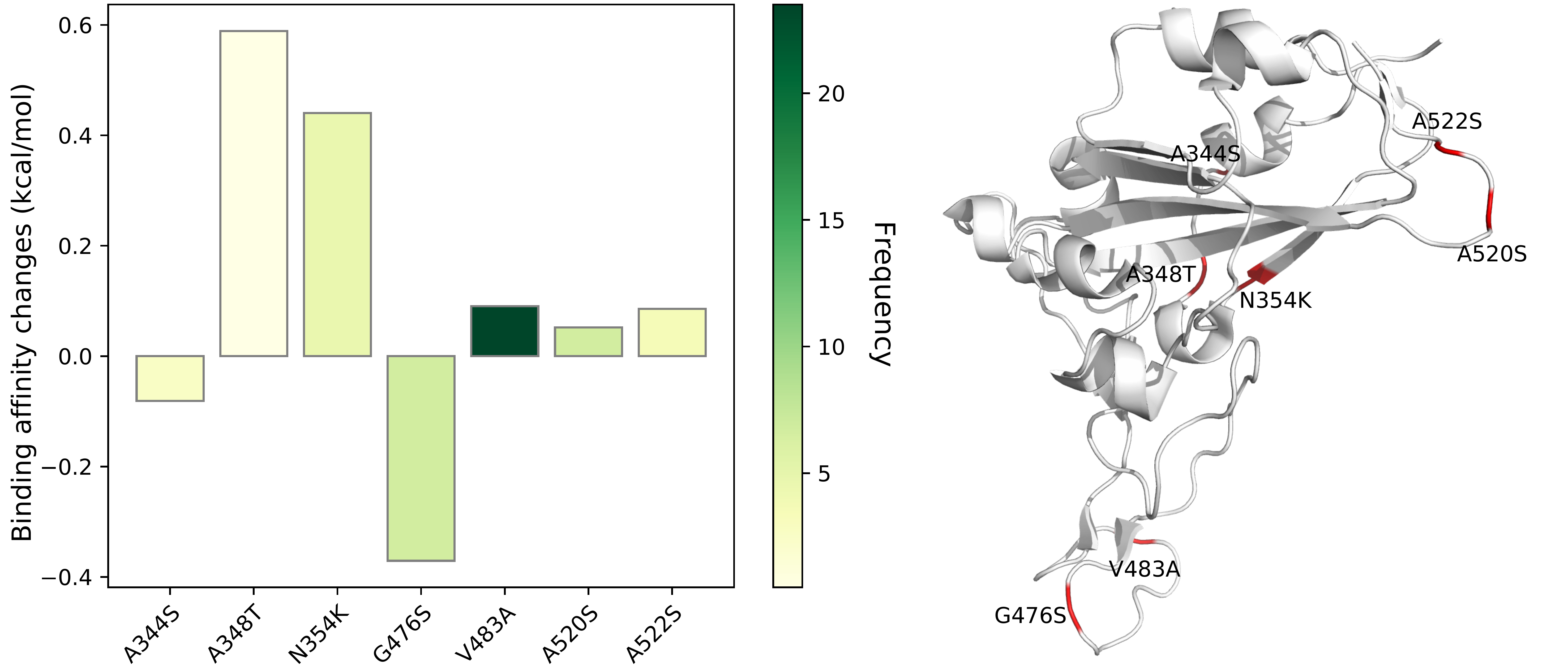}
	\caption{Cluster V. Left: Binding affinity changes $\Delta\Delta G$  induced by   mutations in Cluster V. Right: mutation locations on the SARS-CoV-2 S protein RBD.}
	\label{fig:c5}
\end{figure}

Figure~\ref{fig:c5} presents the binding affinity changes in the fifth cluster. Five of seven mutations have positive free energy changes. Considering the magnitude of the BA change range is within 0.60 kcal/mol, this cluster has a very minor increase in infectivity. Most samples in this cluster were submitted by the US. 

\subsubsection{Cluster VI analysis} 
\begin{figure}
	\includegraphics[width=1\linewidth]{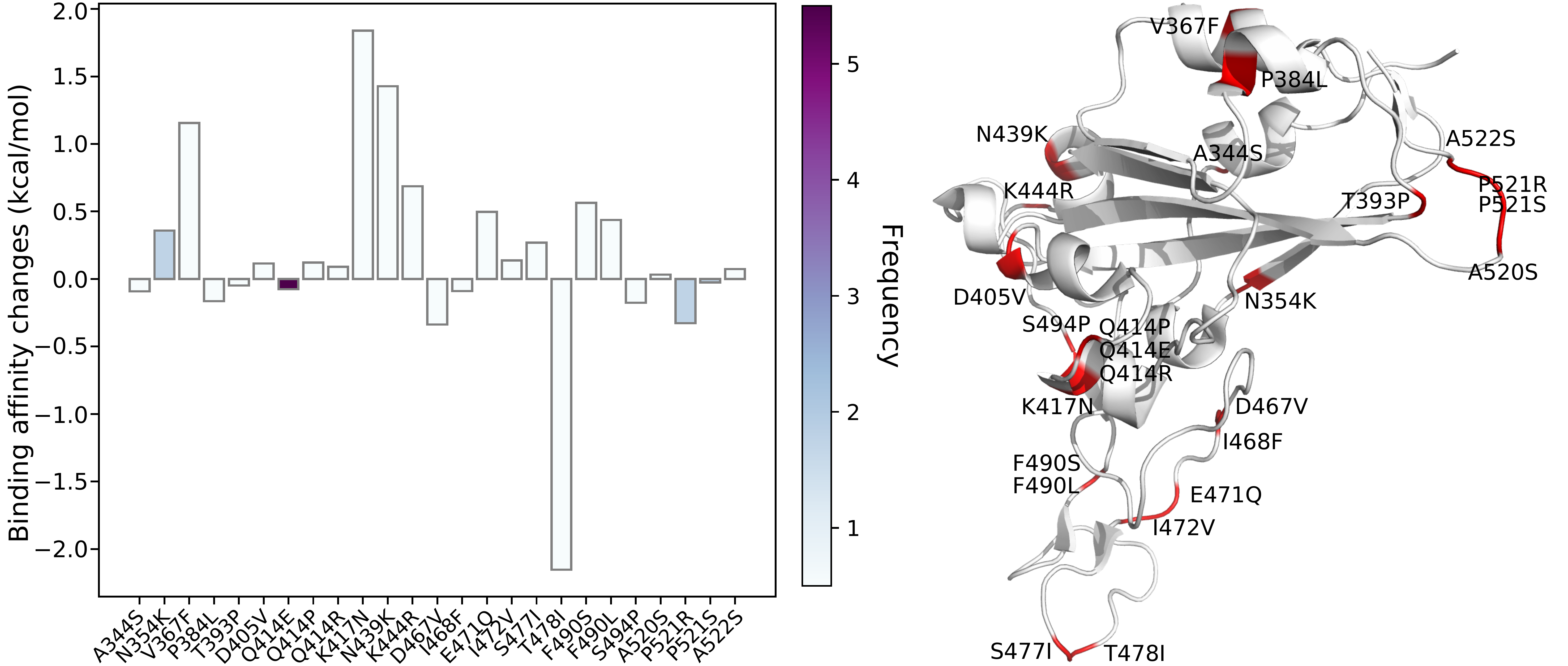}
	\caption{Cluster VI. Left: Binding affinity changes $\Delta\Delta G$  induced by   mutations in Cluster VI. Right: mutation locations on the SARS-CoV-2 S protein RBD.}
	\label{fig:c6}
\end{figure}

The binding affinity changes in the last cluster is shown in Figure~\ref{fig:c6}. Obviously, most mutations on Cluster VI have enhanced the binding affinity of the S protein and ACE2 receptor except T478I. The most significant positive free energy change is caused by mutation K417N. Overall, Cluster VI has strengthened infectivity. This cluster involves genome samples submitted from all countries except for Japan and South Korea. The US submitted most samples. Cluster VI is a new cluster of SARS-CoV-2 according to its low frequency of each mutation.

In summary, five of six clusters, Clusters I, II, IV, V, and VI, have moderate or minor positive binding affinity changes which indicate that the evolution of the SARS-CoV-2 trend to increase infectability by increasing its binding affinity with ACE2 receptor. One cluster, Cluster III, reduced its binding affinity, indicating a decrease in its infectivity. 

\subsection{Impacts of most likely future RBD mutations}

In this section and the next section, we analyze the impacts of all of 3686 possible mutations on 194 S protein RBD of 194 residues. On each amino acid, we classify all  19 possible mutations into most likely future mutations,  likely future mutations, and unlikely future mutations.  Here,  most likely, likely, and unlikely future mutations are defined by the protein mutations induced by only one, simultaneous two, and simultaneous three of genetic changes on three underlying nucleotides on a codon.  Based on the codon analysis of all 194 amino acid residues on the RBD, we have 1149 most likely, 1912 likely, and 625 unlikely mutations. 

\begin{figure}
	\includegraphics[width=1\linewidth]{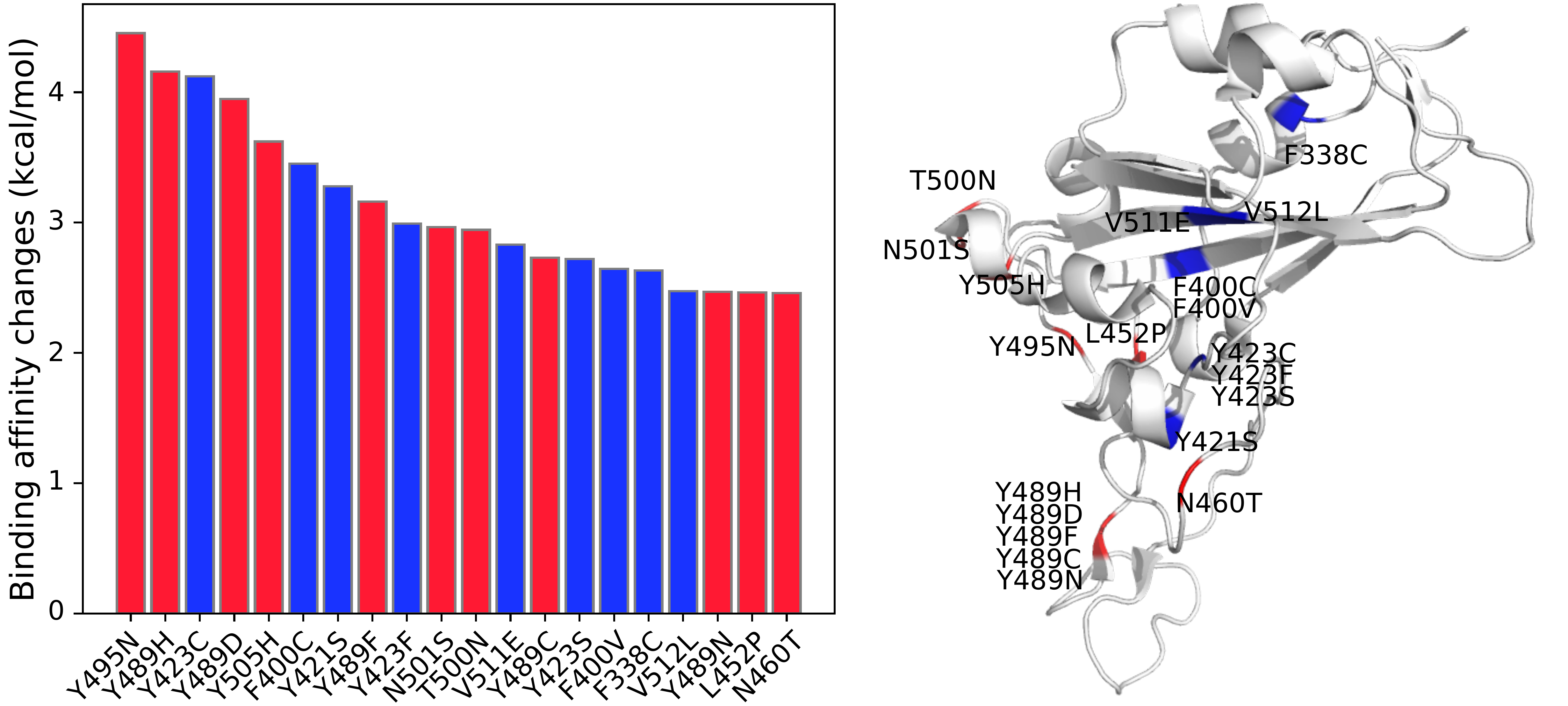}
	\caption{Top 20  most likely future mutations that will strengthen the SARS-CoV-2 infectivity. 
Left:     binding affinity changes $\Delta\Delta G$. Right: mutations on the RBD.  
Red color indicates mutations on the RBM and blue color indicates mutations away from the RBM.
}
	\label{fig:RBM1_top20}
\end{figure}

We compute the $\Delta\Delta G$s following most likely future mutations on the RBD.  
Figure     \ref{fig:RBM1_top20} depicts  20 most likely future mutations that can have the highest adversarial impacts on COVID-19 infectivity.  First, it is noted that mutation Y495N on the RBM has the highest free energy change and if it occurs, it will make the virus significantly more infectious. 
Additionally, mutation Y489H on the RBM would incur another large infectivity strengthening. 
It is worthy to note that residue 489 is a potentially hot spot, where 5 possible mutations, Y489H,  Y489D, Y489F, Y489C, and Y489N,  will lead to the strengthened S protein-ACE2 binding. The other potentially hot spot is residue 423 with Y423C, Y423F, and Y423S being infectivity-strengthening mutations.  Residue 452 on the RBM has been proven to hot spot as it has an existing mutation L452R (see Fig.     \ref{fig:c4}) and another infectivity strengthening mutation,  L452P. 
In general, the highest free energy changes are due to mutations on the RBM. However, mutations away from the RBM can have a considerable impact on the infectivity as well. 

\begin{figure}
	\includegraphics[width=1\linewidth] {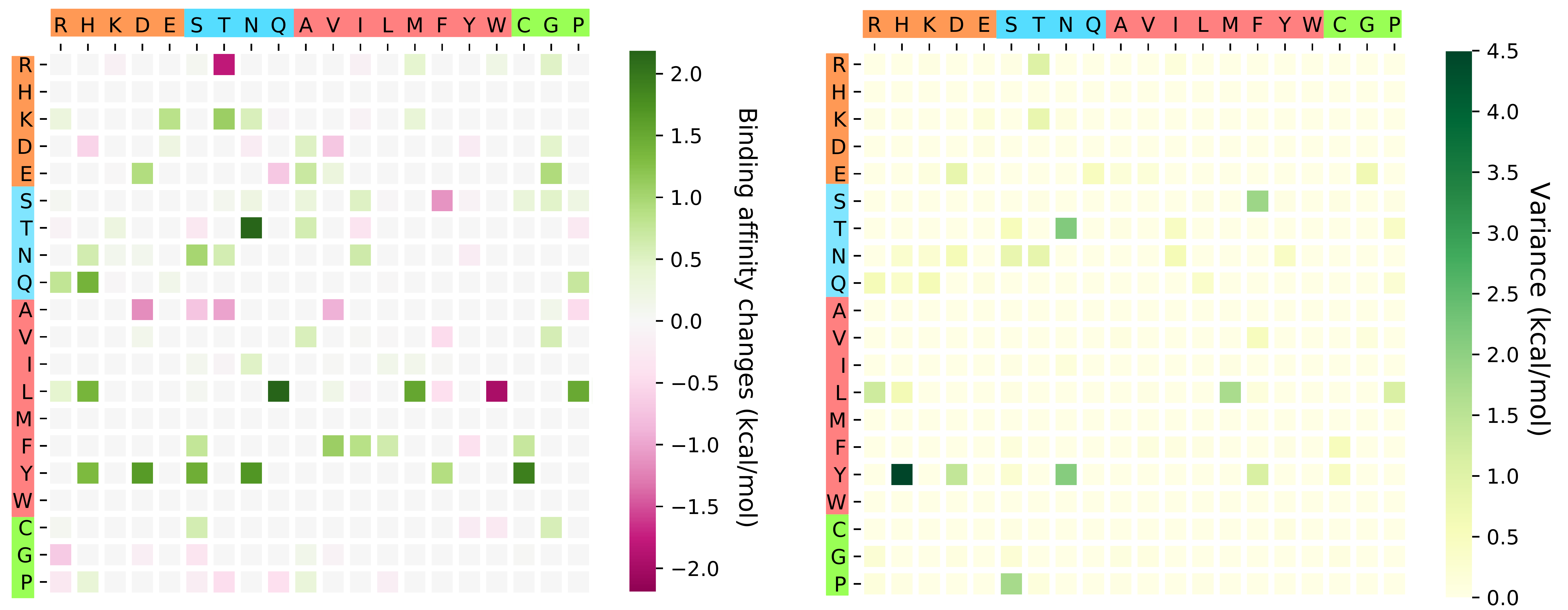}
	\caption{An illustration of the average and variance of $\Delta\Delta G$ (kcal/mol) for most likely mutation types on the RBM. $y$-axes: wild type residues; x-axises: mutant type residues. Colors on the axes indicate residue types.}
	\label{fig:RBM1_avg_var}
\end{figure}
\begin{figure}
	\includegraphics[width=1\linewidth]{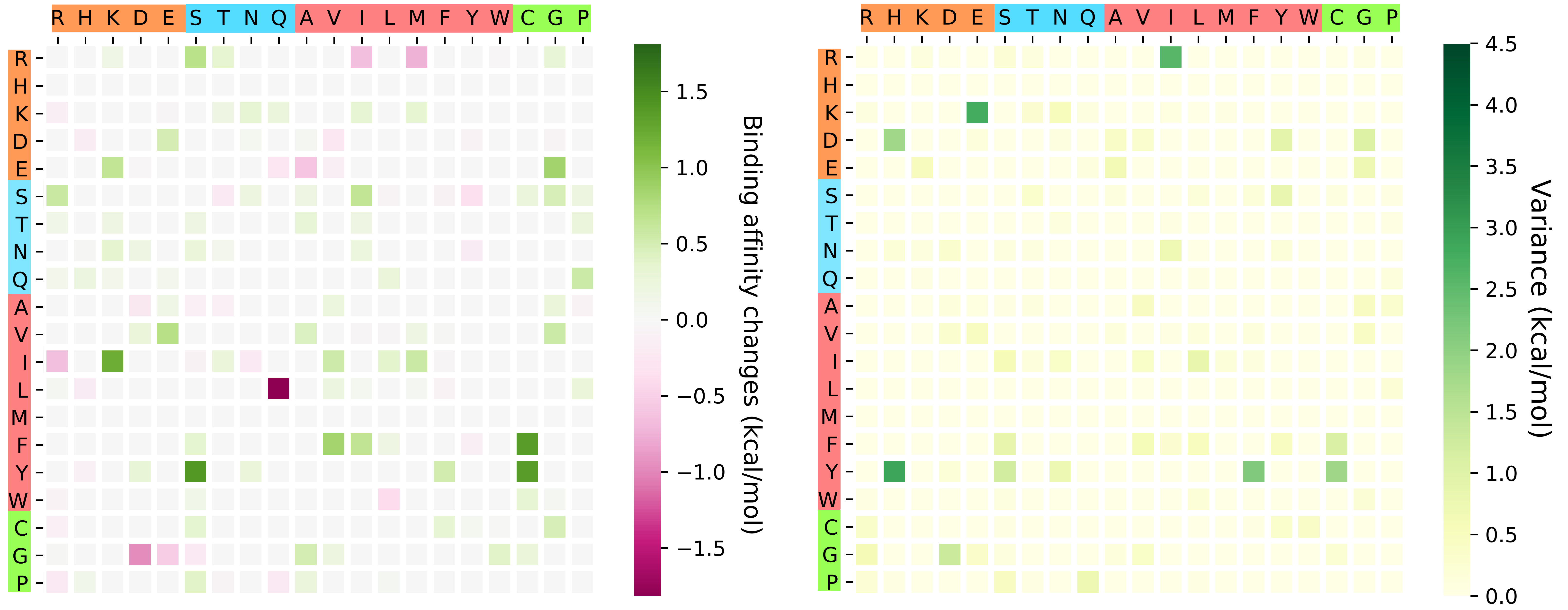}
		\caption{An illustration of the average and variance of $\Delta\Delta G$ (kcal/mol) for most likely mutation types away from the RBM. $y$-axes: wild type residues; x-axes: mutant type residues. Colors on the axes indicate residue types.}
	\label{fig:RBD1_avg_var}
\end{figure}

The above analysis considers only BA strengthening mutations. 
To have a global view of how future mutations would change the COVID-19 infectivity, we analyze the general trend of the free energy changes of most likely mutations according to 400 possible mutation types.   The $\Delta\Delta G$ values following mutations on each amino acid are predicted and averaged by their mutation types. Figure~\ref{fig:RBM1_avg_var} shows the average and variance of $\Delta\Delta G$ (kcal/mol) of each mutation type for most likely mutations on the RBD.  
Here,  $y$-axes stand for wild type residues, and  $x$-axes are mutant type residues. The colors on the axes are the residue types, such as charged, polar uncharged, hydrophobic, and special cases.  
The colors in the heat maps indicate the binding affinity changes strengths and directions.  
It is worthy to note that there are more positive binding affinity changes (green cubes) than negative changes (pink cubes), showing a  trend of more infectious COVID19 strains due to most likely future mutations.  For example, if a wild type mutation takes place from wide type K, T, N, Q, L, F, or Y to any other residue type except for W, it will end up with a more infectious COVID19 strain. However, mutations from R to T,  or from A to many other residue types might lead to a less infectious COVID-19.  The large values on the variance map  indicate where the above average values might not be reliable.   It is seen that the variances are general small. 
Figure \ref{fig:RBD1_avg_var} shows a similar trend for the most likely mutations away from the RDM. 
 
 \subsection{Impacts of   likely and unlikely future RBD mutations}
\begin{figure}
	\includegraphics[width=1\linewidth]{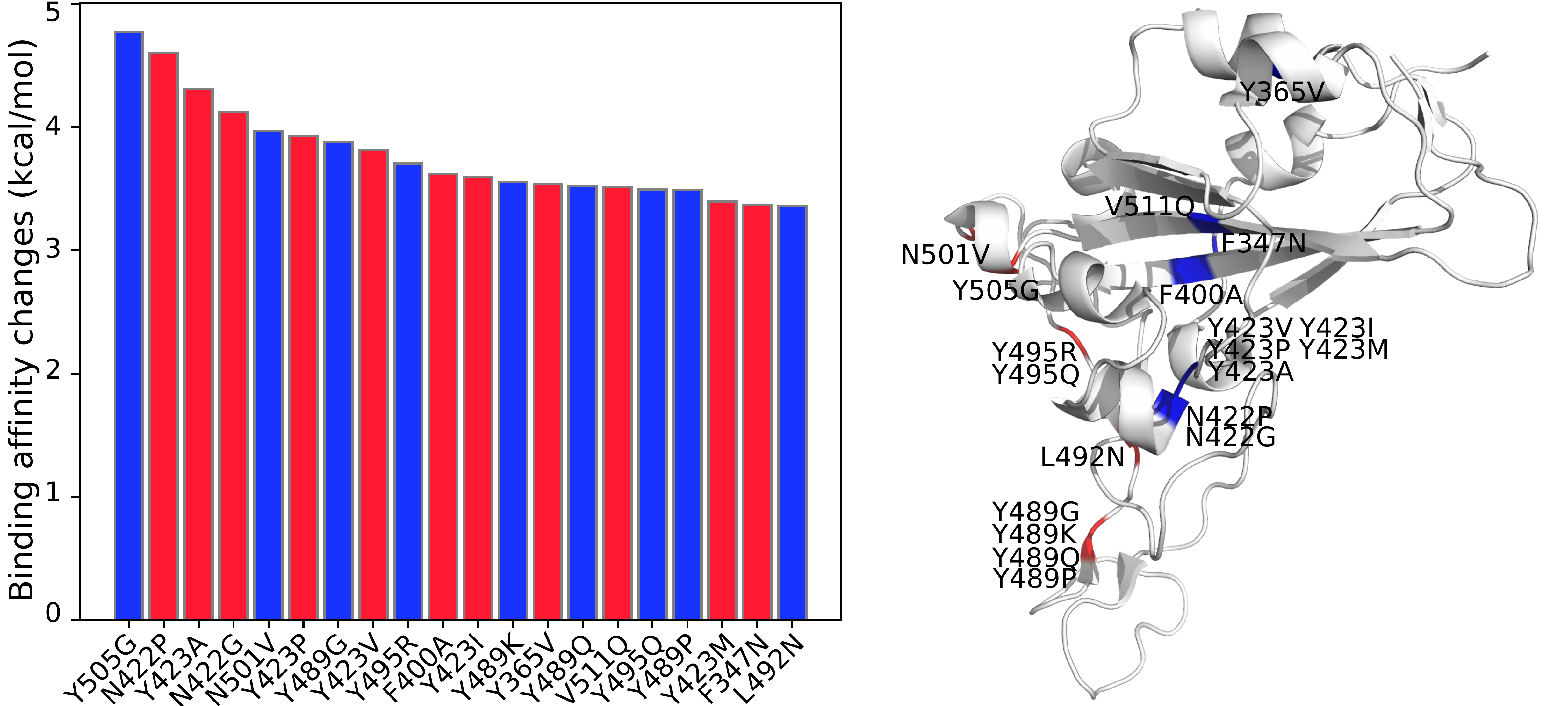}
		\caption{Top 20  likely future mutations that will strengthen the COVID-19 infectivity. 
Left:     binding affinity changes $\Delta\Delta G$. Right: mutations on the RBD.  
Red color indicates mutations on the RBM and blue color indicates mutations away from the RBM.
}
	\label{fig:RBM23_top20}
\end{figure}

\begin{figure}
	\includegraphics[width=1\linewidth]{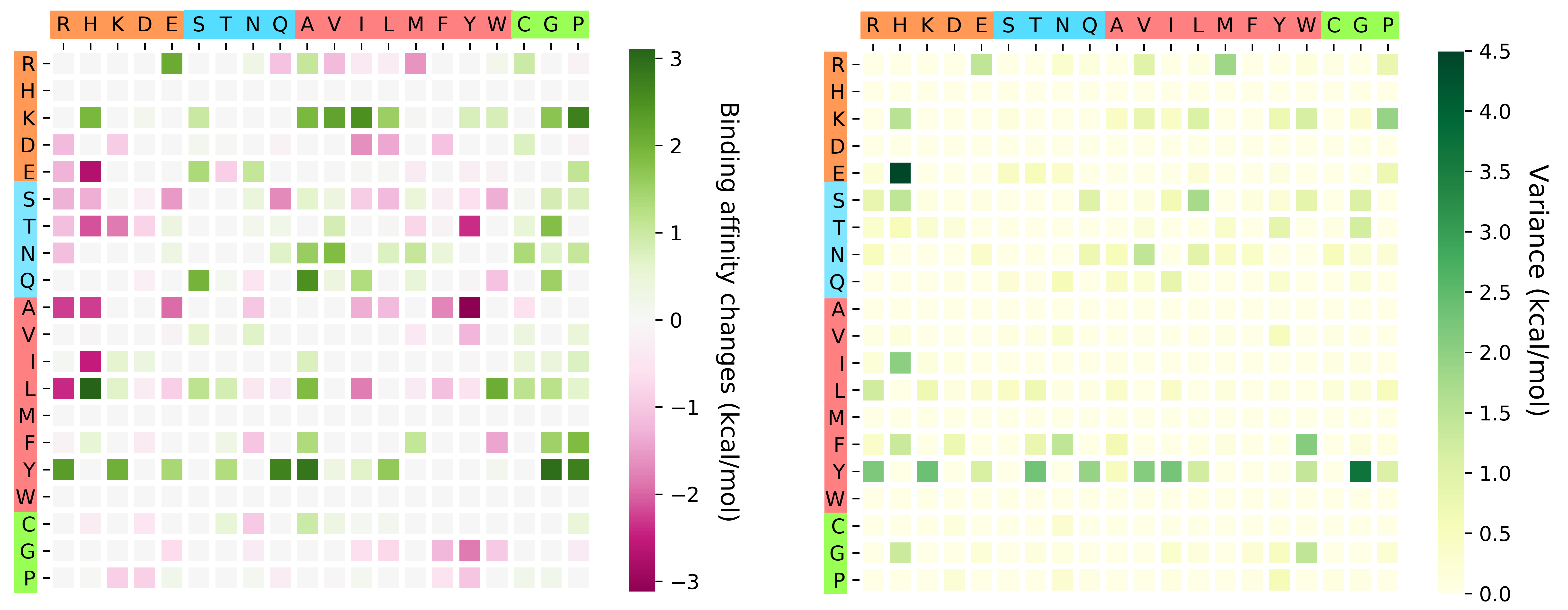}
	\caption{An illustration of the average and variance of $\Delta\Delta G$ (kcal/mol) for likely mutation types on the RBM. $y$-axes: wild type residues; x-axises: mutant type residues. Colors on the axes indicate residue types.}
	\label{fig:RBM2_avg_var}
\end{figure}

\begin{figure}
	\includegraphics[width=1\linewidth]{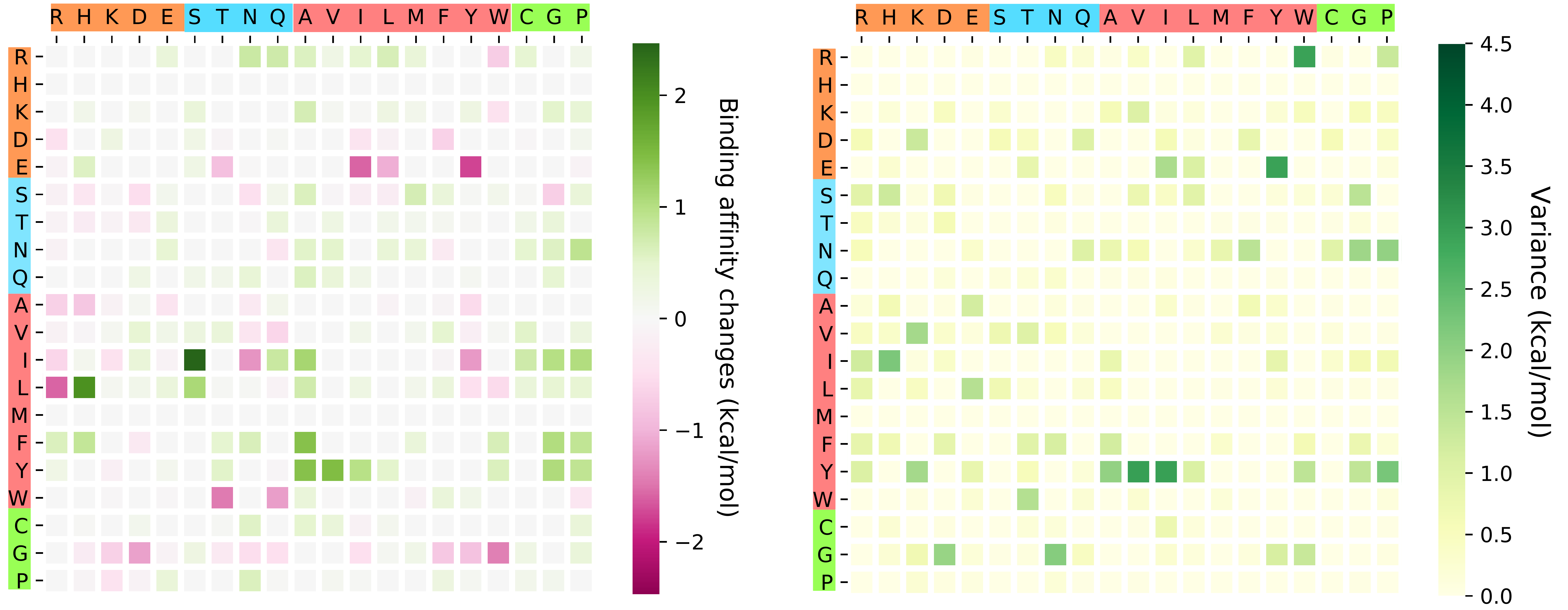}
	\caption{An illustration of the average and variance of $\Delta\Delta G$ (kcal/mol) for likely mutation types away from the RBM. $y$-axes: wild type residues; x-axises: mutant type residues. Colors on the axes indicate residue types.}
	\label{fig:RBD2_avg_var}
\end{figure}

As discussed above, likely and unlikely future mutations require two and three concurrent nucleotide mutations on each codon to happen, respectively.  Figure \ref{fig:RBM23_top20} presents the top 20  likely future mutations that will strengthen the COVID-19 infectivity.  The most energetic adversarial mutation is Y505G.  Note that residue 505 has is a most likely mutation Y505H shown in Fig.      \ref{fig:RBM1_top20}. Therefore, residue 505 is a potentially hot spot on the RBM. The next few energetic adversarial mutations are away from the RBM. Among them, N423P and N422G are hot-spot mutations.  Figure      \ref{fig:RBM1_top20} shows that residue 423 has three most likely energetic mutations while in Fig.  \ref{fig:RBM23_top20},  it has the other three likely energetic mutations. Similarly, residue 489 on the RBM has 5 most likely energetic mutations (see Fig.      \ref{fig:RBM1_top20}) and 4 likely energetic mutations as shown in Fig. \ref{fig:RBM23_top20}. It is on our top surveillance list for the next generation of infectious COVID-19 strains.  Another potentially hot spot is residue 495. 
Figure~\ref{fig:RBM2_avg_var} shows the average and variance of $\Delta\Delta G$ (kcal/mol) of all likely mutations on the RBM where values of most likely mutations and unlikely mutations are excluded. About the same amount of mutations have positive and negative binding affinity changes. In Figure~\ref{fig:RBD2_avg_var}, similar results are shown for the RBD excluding the RBM. Interestingly, mutations on the RBM have larger magnitude changes rather than out of this region for second potential mutations. It again shows that the RBM is the most important region to study. 

\begin{figure}
	\includegraphics[width=1\linewidth]{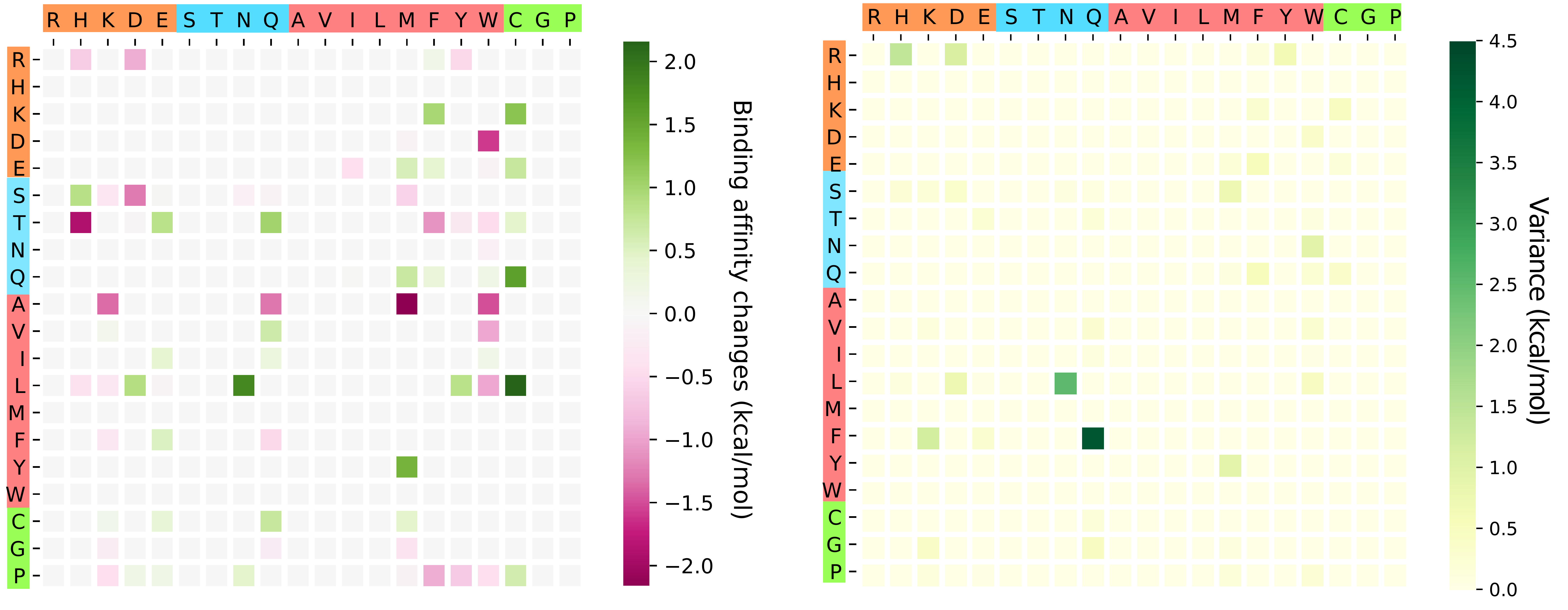}
	\caption{An illustration of the average and variance of $\Delta\Delta G$ (kcal/mol) for unlikely mutation types on the RBM. $y$-axes: wild type residues; x-axises: mutant type residues. Colors on the axes indicate residue types.}
	\label{fig:RBM3_avg_var}
\end{figure}

\begin{figure}
	\includegraphics[width=1\linewidth]{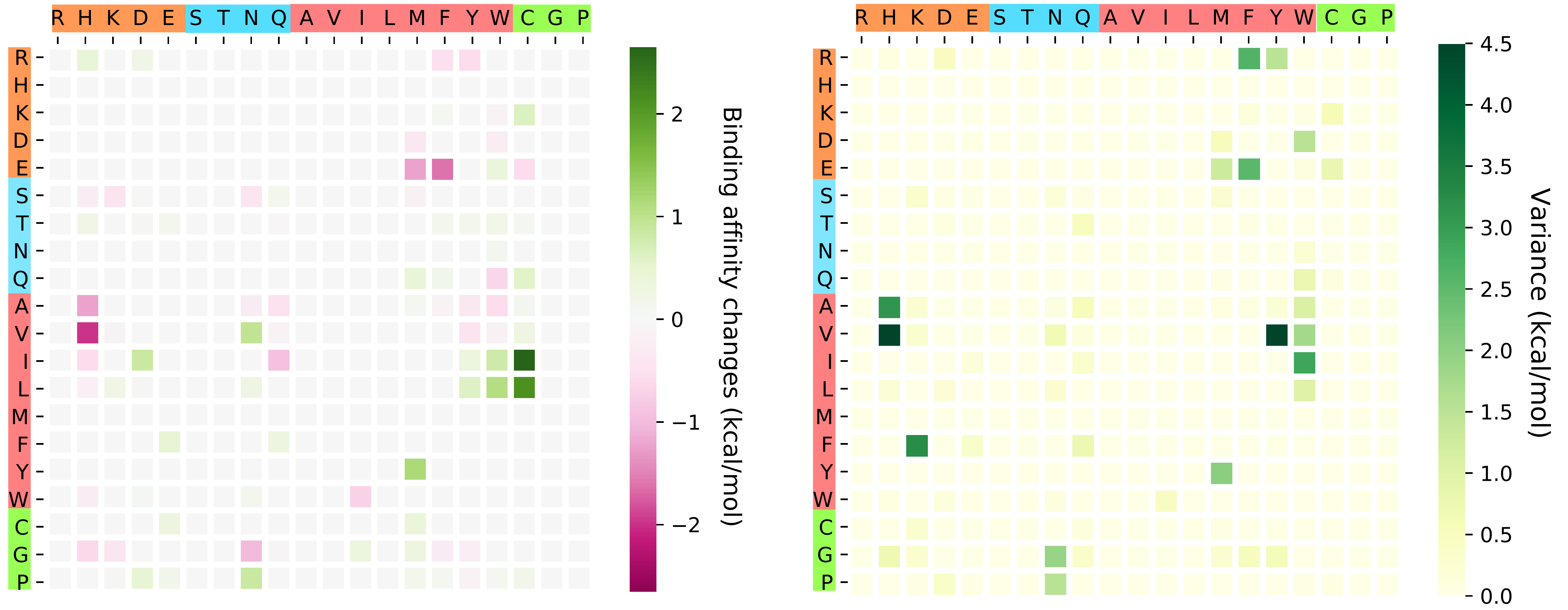}
	\caption{An illustration of the average and variance of $\Delta\Delta G$ (kcal/mol) for most likely mutation types away from the RBM. $y$-axes: wild type residues; x-axises: mutant type residues. Colors on the axes indicate residue types.}
	\label{fig:RBD3_avg_var}
\end{figure}

Figure~\ref{fig:RBM3_avg_var} and \ref{fig:RBD3_avg_var} show the predictions of free energy changes due to unlikely mutations. These mutations have a balanced positive and negative binding affinity changes. We do not expect these mutations to occur in the near future.



\section{Discussion}

\subsection{Conservation analysis via sequence alignment}
\begin{figure}
	\includegraphics[width=1\linewidth]{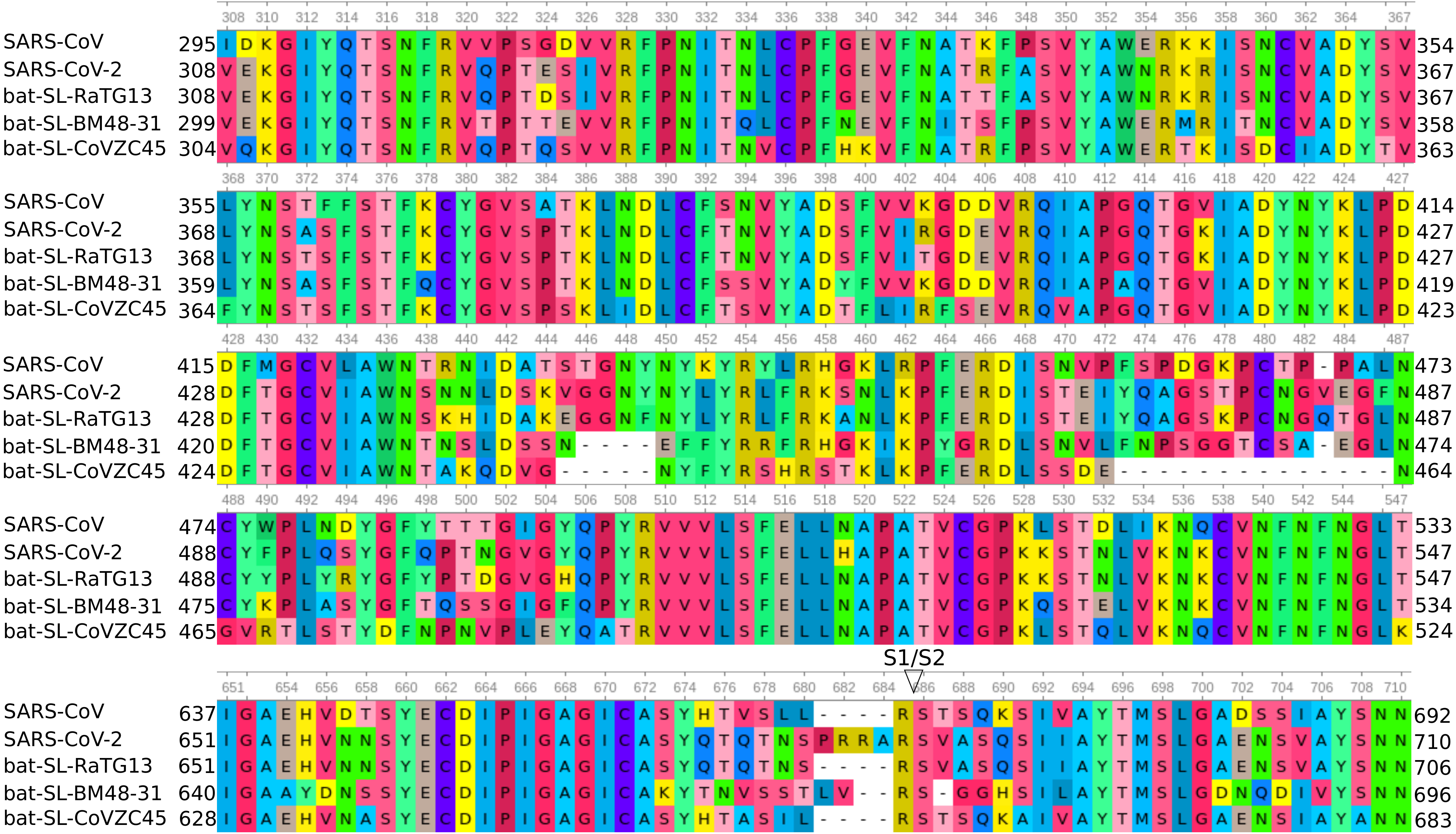}
	\caption{Sequence alignments of  SARS-CoV-2 S protein with those of closely related species, including  SARS-CoV \cite{lee2003major}, bat coronavirus RaTG13 \cite{zhou2020pneumonia}, bat coronavirus BM48-31 \cite{drexler2010genomic}, and bat coronaviruse CoVZC45 \cite{hu2018genomic}. Detailed numbering is given according to SARS-CoV-2. Residue 364 Ala (A) of bat coronavirus BM48-31 is omitted.}
	\label{fig:alignment}
\end{figure}

To further understand the evolution trend and potential infectivity changes of the COVID-19, we carry out S protein sequence alignment analysis to examine residue conservativeness. Figure     \ref{fig:alignment} presents the alignment analysis of SARS-CoV-2 S protein sequence and those of the other four closely related species, namely, SARS-CoV, bat coronavirus RaTG13, bat coronavirus BM48-31, and bat coronavirus CoVZC45. We note that among the residues we discussed in the last section, 414, 422, 423, 492, and 495 are very conservative. They have not undergone any mutations among five related species. In contrast, RBM residues 452, 489, 500, 501, and 505 have a history of mutations and are non-conservative.   Therefore, the predicted infectivity-strengthening mutations on these residues are more likely to happen.      

\subsection{Relative infectivity change analysis for SARS-CoV and SARS-CoV-2}

\begin{figure}
	\includegraphics[width=1\linewidth]{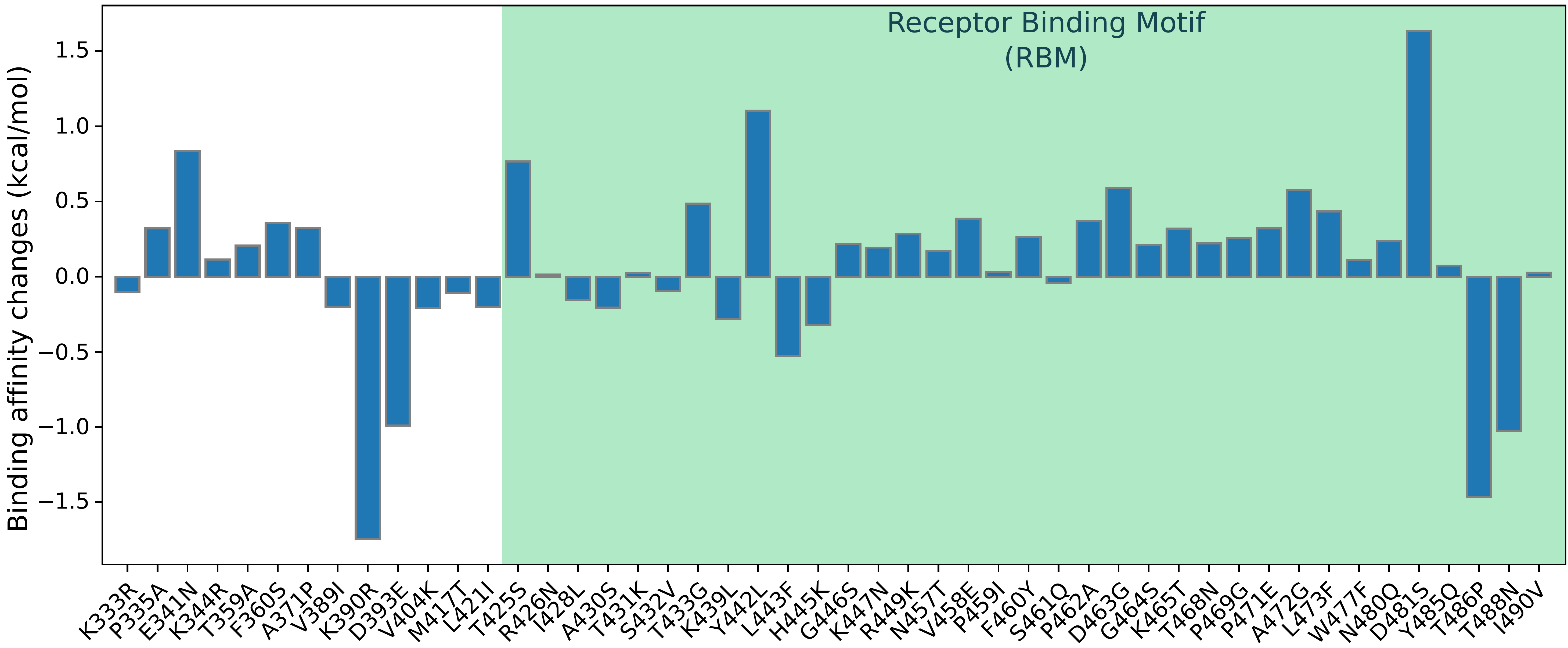}
	\caption{Overall binding affinity changes $\Delta\Delta G$ on the S protein receptor-binding domain (RBD) from SARS-CoV to SARS-CoV-2.  The blue color region marks the binding affinity changes on the receptor-binding motif (RBM). The height of each bar indicates the predicted $\Delta\Delta G$. Residues are labeled according to PDB ID 3D0G \cite{li2008structural}.
 }
	\label{fig:overallSARS-SARS2}
\end{figure}

As mentioned earlier, there are inconsistent assessments about relative infectivity between SARS-CoV
and SARS-CoV-2 in the literature \cite{wrapp2020cryo, shang2020structural,walls2020structure}.  Our validated computational method can be employed to resolve this discrepancy.  Based on the sequence alignment shown in Figure     \ref{fig:alignment}, we conduct binding affinity change calculations for all relevant mutations on the SARS-CoV S protein RBD. Figure     \ref{fig:overallSARS-SARS2} illustrates the S protein-ACE2 binding affinity changes following the mutations from SARS-CoV to SARS-CoV-2. The SARS-CoV S protein and ACE2 complex 3D0G  \cite{li2008structural} is used as the wide type in our predictions. It is interesting to note that overall, there are more infectivity-strengthening mutations than infectivity-weakening mutations on the RBD. This is particularly true for mutations on the RBM. This result indicates that the SARS-CoV-2 sample collected on January 5, 2020, \cite{wu2020new} is slightly more infectious than SARS-CoV found in 2003  \cite{lee2003major}. 

For a comparison between various SARS-CoV-2 subtypes and SARS-CoV of 2003, our results indicate that SARS-CoV-2 Clusters, I, II, IV, V, and VI become more infectious than SARS-CoV whereas SARS-CoV-2 Cluster III may have a similar infective rate with that of SARS-CoV. 
 
Compared with SARS-CoV, SARS-CoV-2 has four extra residues, PRRA from 681 to 684, as shown in  Figure     \ref{fig:alignment}.  It is believed that these extra residues might change SARS-CoV-2's behavior in ACE2 assisted entry of host cells  \cite{walls2020structure}.  However, this speculation has no qualitative nor quantitative validation at present.

\section{Material and methods}

\subsection{Sequences and structures}
Amino acid sequences and mutant data of the S protein used in the analysis were obtained from NCBI GenBank and GISAID \cite{shu2017gisaid}. After the first complete genome sequence of SARS-CoV-2 released on NCBI GenBank (Access number: NC\_045512.2) \cite{wu2020new}, there has been a large number of genome sequences. Other sequences from GenBank are as follows: bat coronavirus RaTG13 (MN996532.1)\cite{zhou2020pneumonia}, bat coronavirus BM48-31 (NC\_014470.1) \cite{drexler2010genomic} and bat coronavirus CoVZC45 (MG772933.1) \cite{hu2018genomic}. The mutant information of  13752  whole-genome sequences of S protein with high coverage of SARS-CoV-2 strains from the infected individuals around the world was obtained from the GISAID database \cite{shu2017gisaid} (https://www.gisaid.org/). Data without the exact submission date in GISAID were not considered.

Beta-CoV S protein structures were obtained from the RCSB Protein Data Bank: SARS-CoV RBD with ACE2 (PDB 3D0G) \cite{li2008structural} and  SARS-CoV-2 RBD with ACE2 (PDB 6M0J) \cite{lan2020structure}. The structures were presented by using PyMOL \cite{PyMOL}. Sequences alignments were performed on SARS-CoV-2 S protein sequences by using MAFFT v7.388 \cite{katoh2013mafft} and on SARS-CoV-2 genome by using the Clustal Omega multiple sequence alignment with default parameters \cite{sievers2014clustal}.

\subsection{TopNetTree model for protein-protein interaction (PPI) binding affinity changes upon mutation}
\begin{figure}
\begin{center}
\includegraphics[width=0.88\linewidth]{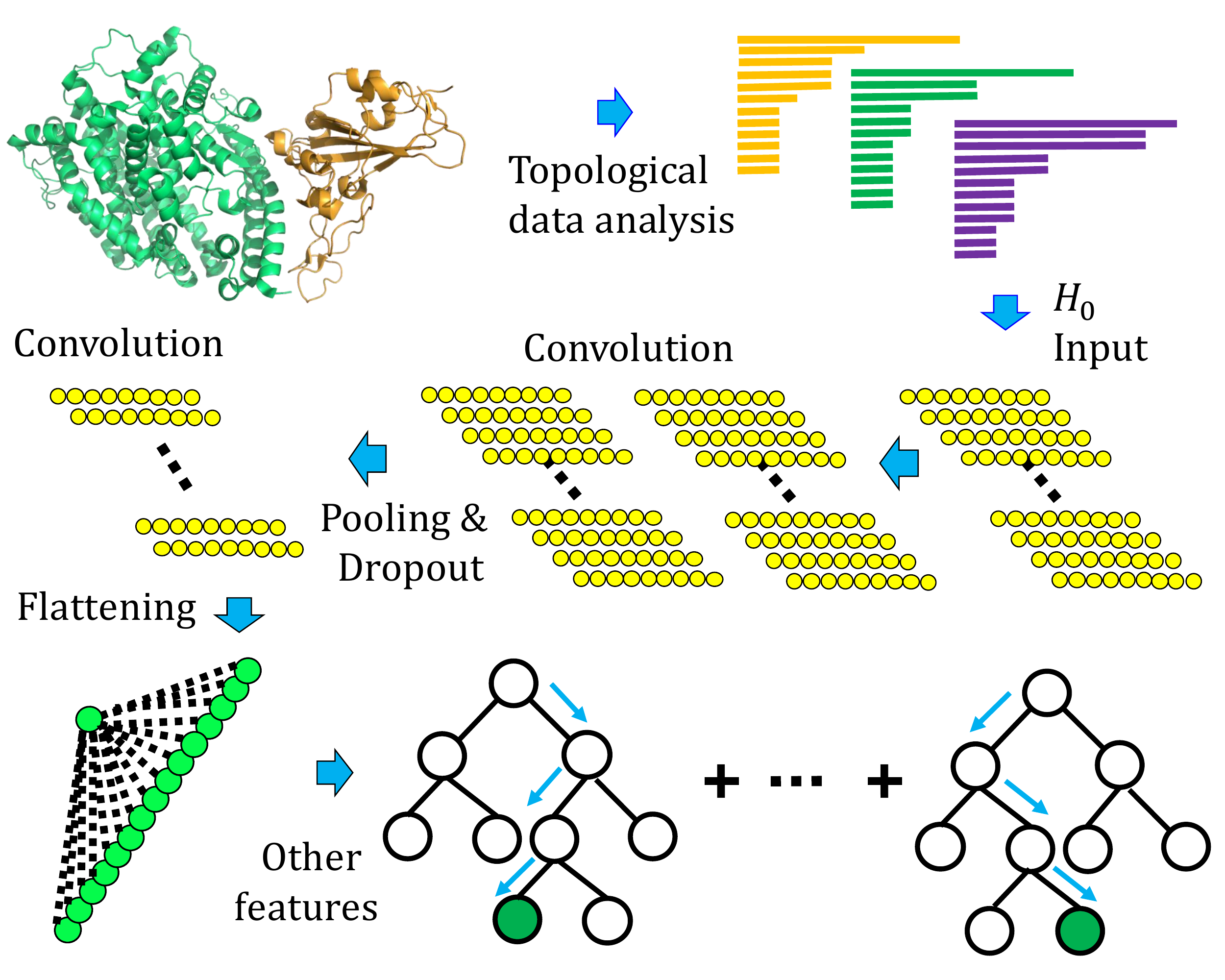}
\end{center}
\caption{An illustration of the TopNetTree model. Protein structure shown in the plot is SARS-CoV-2 spike receptor-binding domain bound with ACE2 (PDB 6M0J). Here, $H_0$ are the 0-dimensional topological  input features for machine learning model.}
\label{fig:NetTree}
\end{figure}

 The topology-based network tree (TopNetTree) is constructed by an innovative integration between the topological representation and NetTree for predicting protein-protein interaction binding affinity changes following mutation $\Delta\Delta G$. In this work, TopNetTree is applied to predict the binding affinity changes of mutations that happened on RBD with ACE2 of SARS-CoV-2 after January 5, 2020. As shown in Fig.~\ref{fig:NetTree}, topology-based feature generation is the first step followed by a convolutional neural network (CNN)-assisted model. The topological representation uses element- and site-specific persistent homology to simplify the structural complexity of protein-protein complexes and encode vital biological information into topological invariants. NetTree is a recently developed deep learning algorithm that integrates the advantages of convolutional neural networks   and gradient-boosting trees (GBT). In this section, we briefly describe the topology representation for machine learning training and prediction. Details can be found in the literature \cite{wang2020topology}.

\subsubsection{Topology-based feature generation of PPIs}

 The topology-based feature generation is built upon from persistence homology starting with simplicial complex and filtration. As a type of algebraic topology, persistence homology studies simplicial complex on discrete datasets under various settings. Among the many constructions, two that are widely used for point clouds are the Vietoris-Rips (VR) complex and alpha complex \cite{edelsbrunner2000topological} which are applied in our approach. Built upon a simplicial complex, the topological invariants of a point-cloud dataset can be identified, such as the set of atoms in protein-protein interactions. Meanwhile, topological invariants (separated components, rings, and cavities) can be enumerated by counting the numbers referred to as Betti-0, Betti-1, and Betti-2, respectively. Thus taxing and uninformative features or calculations are fully abandoned, whereas geometric and topological characteristics persevere as data representation. Moreover, using persistent homology, the original 3D point-cloud data are characterized by topological barcodes that record the ``birth'' and ``death'' of each topological invariants and simplifying complicated structural representations of a PPI-complex. Although topology data presentation much simplifies the problem in many directions, better construction is required for it to extract patterns of different biological or chemical aspects. Before talking about more detailed feature generations, we first preset the constructions for a PPI complex into various subsets. 
\begin{enumerate}
	\item $\mathcal{A}_\text{m}$: atoms of the mutation sites.
	\item $\mathcal{A}_\text{mn}(r)$: atoms in the neighbourhood of the mutation site within a cut-off distance $r$.
	\item $\mathcal{A}_\text{Ab}(r)$: antibody atoms within $r$ of the binding site.
	\item $\mathcal{A}_\text{Ag}(r)$: antigen atoms within $r$ of the binding site.
	\item $\mathcal{A}_\text{ele}(\text{E})$: atoms in the system that has atoms of element type E. The distance matrix is specially designed such that it excludes the interactions between the atoms form the same set. For interactions between atoms $a_i$ and $a_j$ in set $\mathcal{A}$ and/or set $\mathcal{B}$, the modified distance is defined as
	\begin{equation}
	D_{\text{mod}}(a_i, a_j) =
	\begin{cases}
	\infty, \text{ if } a_i, a_j\in\mathcal{A}\text{, or }a_i, a_j\in \mathcal{B}, \\
	D_e(a_i, a_j), \text{ if } a_i\in\mathcal{A} \text{ and }a_j\in \mathcal{B},
	\end{cases}
	\label{eq:modified_equation}
	\end{equation}
	where $D_e(a_i, a_j)$ is the Euclidian distance between $a_i$ and $a_j$.
\end{enumerate}

In algebraic topology, molecular atoms can be treated as points presented by $v_0$, $v_1$, $v_2$, $...$, $v_k$ as $k\!+\!1$ affinely independent points. Simplicial complex, the essential building blocks, is a finite collection of sets of points $K=\{\sigma_i\}$, and $\sigma_i$ are called linear combinations of these points in $\mathbb{R}^n$ ($n\ge k$).
For instance, a 0-, 1-, 2-, or 3-simplex in geometry representation is a vertex, an edge, a triangle, or a tetrahedron, respectively. A simplicial complex $K$ is valied if a face $\tau$ of a $k$-simplex $\sigma_i$ of $K$ is also in $K$, such that $\tau \subseteq \sigma_i$ and $\sigma_i\in K$ imply $\tau \in K$ and the non-empty intersection of any two simplieces is a face for both. Given a simplicial complex $K$, a $k$-chain is a finite formal sum of $k$-simplices; that is, $\sum_{i}\alpha_i\sigma^k_i$. The set of all $k$-chains of the simplicial complex $K$ equipped with an algebraic field (typically, $\mathbb{Z}_2$) forms an abelian group $C_k(K, \mathbb{Z}_2)$, which means the coefficients $a_i$ are chosen from $\mathbb{Z}_2$. A boundary operator $\partial_k: C_k\!\rightarrow\!C_{k-1}$ for a $k$-simplex $\sigma^k=\{v_0,v_1,v_2,\cdots,v_k\}$ are homomorphisms defined as $\partial_k \sigma^k = \sum^{k}_{i=0} (-1)^i\{ v_0, v_1, \cdots, \hat{v_i}, \cdots, v_k \}$, where $\{v_0, v_1, \cdots ,\hat{v_i}, \cdots, v_k\}$ is a $(k\!-\!1)$-simplex excluding $v_i$ from the vertex set. Consequently, a important property of boundary operator, $\partial_{k-1}\partial_k= \emptyset$, follows from that boundaries are boundaryless. Moreover the $k$th cycle group $Z_k={\rm ker} \partial_k=\{c\in C_k \mid \partial_k c=\emptyset\}$ is deffined to be the kernel of $\partial_k$, whose elements are called $k$-cycles; and the $k$th boundary group is the image of $\partial_{k+1}$ denoted as $B_k={\rm im} ~\partial_{k+1}= \{ \partial_{k+1} c \mid c\in C_{k+1}\}$. The algebraic construction to connect a sequence of complexes by boundary maps is called a chain complex
\[
\cdots \stackrel{\partial_{i+1}}\longrightarrow C_i(X) \stackrel{\partial_i}\longrightarrow C_{i-1}(X) \stackrel{\partial_{i-1}}\longrightarrow \cdots \stackrel{\partial_2} \longrightarrow C_{1}(X) \stackrel{\partial_{1}}\longrightarrow C_0(X) \stackrel{\partial_0} \longrightarrow 0
\]
and the $k$th homology group is the quotient group defined by
\begin{equation}
H_k = Z_k / B_k.
\end{equation}
The key property of boundary operators implies $B_k\subseteq Z_k \subseteq C_k$. The Betti numbers are defined by the ranks of $k$th homology group $H_k$ which counts $k$-dimensional holes, especially, $\beta_0\!=\!{\rm rank}(H_0)$ reflects the number of connected components, $\beta_1\!=\!{\rm rank}(H_1)$ reflects the number of loops, and $\beta_2\!=\!{\rm rank}(H_2)$ reveals the number of voids or cavities. Together, the set of Betti numbers $\{\beta_0,\beta_1,\beta_2,\cdots \}$ indicates the intrinsic topological property of a system. 

Persistent homology is devised to track the multiscale topological information over different scales along a filtration \cite{edelsbrunner2000topological}. A filtration of a topology space $K$ is a nested sequence of subspaces $\{K^t\}_{t=0,...,m}$ of $K$ such that $
\emptyset = K^0 \subseteq K^1 \subseteq K^2 \subseteq \cdots \subseteq K^m = K$. Moreover, on this complex sequence, we obtain a sequence of chain complexes by homomorphisms: $C_*(K^0) \to C_*(K^1) \to \cdots \to C_*(K^m)$ and a homology sequence: $H_*(K^0) \to H_*(K^1) \to \cdots \to H_*(K^m)$, correspondingly. The $p$-persistent $k$th homology group of $K^t$ is defined as $H_k^{t,p} = Z^t_k/(B_k^{t+p}\bigcap Z^t_k)$, where $B_k^{t+p} = {\rm im} \partial_{k+1}(K^{t+p})$. Intuitively, this homology group records the homology classes of $K^t$ that are persistent at least until $K^{t+p}$. Under the filtration process, the persistent homology barcodes can be generated. Then the feature vectors can be constructed from these sets of intervals for machine learning models.

  In a variety of vectorization methods, one discretizes the filtration parameter interval into bins and model the behavior of the barcodes in each bin \cite{cang2018representability}. To make use of advanced machine learning algorithms, we subdivide a filtration interval into bins of length. Then the numbers of persistence intervals are counted for each bin, such that birth events and death events can be represented. This approach gives us three feature vectors for each topological barcode for the machine learning method. Note for different discretizations, the characterization of birth and death might not be stable so that only Betti-0 ($H_0$) barcodes obtained from the VR filtration are applied in this approach. Intuitively, features generated by binned barcode vectorization can reflect the strength of atom bonds, van der Waals interactions, and can be easily incorporated into a CNN, which captures and discriminates local patterns. Another method of vectorization is to get the statistics of bar lengths, birth values, and death values, such as sum, maximum, minimum, mean, and standard derivation. This method is applied to vectorize Betti-1 ($H_1$) and Betti-2 ($H_2$) barcodes obtained from alpha complex filtration based on the facts that higher-dimensional barcodes are sparser than $H_0$ barcodes.

\subsubsection{Machine learning models}
Prediction of binding affinity changes following mutation for PPIs is very challenging due to the complex dataset and 3D structures. A hybrid machine learning algorithm that integrates a CNN and GBT is designed to overcome difficulties. Briefly speaking, partial topologically simplified descriptions, specifically vectorized $H_0$ barcode feature, are converted into concise features by the CNN module. Then a GBT module is trained on the whole feature set for a robust predictor with effective control of overfitting.

{\bf TopGBT model}. The gradient boosting tree (GBT) method produces a prediction model as an ensemble method which is a class of machine learning algorithms. It builds a powerful module for regression and classification problems from weak learners. By the assumption that the individual learners are likely to make different mistakes, the method using a summation of the weak learners to eliminate the overall error. Furthermore, a decision tree is added to the ensemble depending on the current prediction error on the training dataset. Thus this method (a topology-based GBT or TopGBT) is relatively robust against hyperparameter tuning and overfitting, especially for a moderate number of features.  The GBT is shown for its robustness against overfitting, good performance for moderately small data sizes, and model interpretability. The current work uses the package provided by scikit-learn (v 0.23.0).

{\bf TopCNN model}. CNN is a class of deep neural networks and is considered as the most successful architectures. CNN is a regularized case of a multilayer connected neural network, such that each neuron is connected locally to neurons in the next convolution layers and the weights are shared across different locations. To prepare the integration of CNN and GBT, CNN is treated as an intermediate model that converts vectorized $H_0$ features into a higher-level abstract feature for the downstream model.

{\bf TopNetTree model}. A supervised CNN model with the PPI $\Delta\Delta G$ as labels is trained for extracting high-level features from $H_0$ barcodes. Once the model is set up, the flatten layer neural outputs of CNN are feed into a GBT model to rank their importance. Based on the importance, and ordered subset of CNN-trained features is combined with features constructed from high-dimensional topological barcodes, $H_1$ and $H_2$ into the final GBT model as shown in Fig.~\ref{fig:NetTree}. As for the parameters of the GBT model, 10 times 10-fold experiments are done for searching the optimal parameter setting. 

\subsubsection{Cross-validation of TopNetTree}
\begin{table}[!htb]
	\caption{Ten-fold cross-validation of the TopNetTree on the SKEMPI 2.0 dataset.  }
	\centering
	
	\begin{tabular}{lccc|lccc}
		\toprule
		& $R_p$ & $\tau$ & RMSE (kcal/mol) & & $R_p$ & $\tau$ & RMSE (kcal/mol)\\
		\midrule
		Fold 1 (Train) & 0.981 & 0.884 & 0.366 & Fold 6 (Train) & 0.983 & 0.904 & 0.353 \\
		Fold 1 (Test)  & 0.835 & 0.595 & 1.065 & Fold 6 (Test)  & 0.836 & 0.594 & 1.064 \\
		\midrule
		Fold 2 (Train) & 0.982 & 0.902 & 0.360 & Fold 7 (Train) & 0.983 & 0.904 & 0.356 \\
		Fold 2 (Test)  & 0.839 & 0.600 & 1.061 & Fold 7 (Test)  & 0.838 & 0.594 & 1.060\\
		\midrule
		Fold 3 (Train) & 0.982 & 0.887 & 0.366 & Fold 8 (Train) & 0.979 & 0.878 & 0.392 \\
		Fold 3 (Test)  & 0.837 & 0.595 & 1.068 & Fold 8 (Test)  & 0.840 & 0.596 & 1.061 \\
		\midrule
		Fold 4 (Train) & 0.981 & 0.880 & 0.369 & Fold 9 (Train) & 0.982 & 0.902 & 0.362 \\
		Fold 4 (Test)  & 0.841 & 0.596 & 1.059 & Fold 9 (Test)  & 0.838 & 0.596 & 1.069\\
		\midrule
		Fold 5 (Train) & 0.982 & 0.906 & 0.365 & Fold 10 (Train)& 0.982 & 0.886 & 0.367 \\
		Fold 5 (Test)  & 0.839 & 0.594 & 1.062 & Fold 10 (Test) & 0.835 & 0.596 & 1.064 \\
		\midrule
		Average (Train) & 0.982 & 0.893 & 0.366 & & & &  \\
		Average (Test)  & 0.838 & 0.596 & 1.063 & & & & \\
		\bottomrule
	\end{tabular}
	\label{tab:10-fold}
\end{table}

The proposed TopNetTree method is trained on the SKEMPI 2.0 dataset \cite{jankauskaite2019skempi}, which has 4,169 variants in 319 different complexes. A set ``S8338'' with 8,338 variants was derived from SKEMPI 2.0 dataset by setting the reverse mutation energy changes to the negative values of its original energy changes. To address the reliability of the TopNetTree method, we did the tenfold cross-validation on the SKEMPI 2.0 dataset with the averaged training accuracy, Pearson correlation coefficients $R_\text{p}$, Kendall's $\tau$, and the root mean square error (RMSE), being 0.98, 0.89, and 0.37 kcal/mol. As shown in Table~\ref{tab:10-fold}, these metrics are based on the average of ten ten-fold cross-validations which indicate TopNetTree is well trained. The performance test of tenfold cross-validation on dataset gives as $R_\text{p}=0.84$, $\tau = 0.60$, and RMSE $=1.06$ kcal/mol, which is of the same level of accuracy as the best in the literature \cite{wang2020topology}.

\section{Conclusion}
The infectivity of severe acute respiratory syndrome coronavirus 2 (SARS-CoV-2) is a vital factor for preventive measurements against coronavirus disease 2019 (COVID-19) and reopening the global economy \cite{wrapp2020cryo, shang2020structural,walls2020structure}. However, it is very challenging to rigorously determine the viral infectivity experimentally and quantitatively compare the relative infectivity between SARS-CoV and SARS-CoV-2. These challenges are deteriorated by the continuous evolution of  SARS-CoV-2 due to its existing 13752 SNP variants in six distinct clusters \cite{wang2020decoding}. In the present work, we develop an advanced TopNetTree method based on algebraic topology and deep learning to predict the spike glycoprotein (S protein) and the host angiotensin-converting enzyme 2 (ACE2) binding affinity changes induced by mutations. Based on binding affinity changes, we reveal that mutations have made five out of six clusters of SARS-CoV-2 more infectious than the original virus found in Wuhan \cite{wu2020new}. Additionally, based on sequence alignment and mutation-induced binding affinity changes, we show that SARS-CoV-2 \cite{wu2020new} is slightly more infectious than SARS-CoV found in 2003 \cite{lee2003major}. Finally, we systematically compute the binding affinity changes of all possible 3686 future mutations to unveil that the most likely mutations will further strengthen  SARS-CoV infectivity.  
We predict that residues 452, 489, 500, 501, and 505 on the receptor-binding motif (RBM) have high chances to mutate into significantly more infectious COVID-19 strains.   

\section*{Supporting Material}
Supporting Material is available for tables of (1) Six clusters of SARS-CoV-2 mutations on the RBD and predicted BA changes; (2), (3), and (4) Predicted BA changes for most likely, likely, and unlikely future mutations, respectively; and (5) Predicted BA changes for all mutations from SARS-CoV to SARS-CoV-2.

\section*{Acknowledgment}
This work was supported in part by NIH grant  GM126189,   NSF Grants DMS-1721024,  DMS-1761320, and IIS1900473,  Michigan Economic Development Corporation,  Bristol-Myers Squibb,  and Pfizer.
The authors thank The IBM TJ Watson Research Center, The COVID-19 High Performance Computing Consortium, NVIDIA, and MSU HPPC for computational assistance. RW  thanks Dr. Changchuan Yin for assistance. 


\end{document}